\documentclass[conference]{IEEEtran}

\newcommand{\ignore}[1]{}

\usepackage{epsfig}
\usepackage{xspace}
\usepackage{color}
\usepackage{xcolor}
\usepackage{ifthen}
\usepackage{fancyvrb}
\usepackage{booktabs}
\usepackage{amssymb}
\usepackage{subcaption}
\usepackage{listings}
\usepackage{algorithm2e}
\usepackage{float}
\usepackage{mparhack}

\IEEEsettopmargin{t}{.875in}
\IEEEsettextheight{.875in}{.875in}
\usepackage{amsmath}
\usepackage{amsthm}
\usepackage{amsbsy}

\newcommand{\TITLE}{Poor Video Streaming Performance Explained (and Fixed)}

\newcommand{\COLOR}{no}
\newcommand{\PAGENUMBERS}{yes}
\newcommand{\SHOWTOAPPEAR}{no}
\newcommand{\COMMENTS}{no}

\newcommand{\system}{Sprint\xspace}

\newcommand{\systemx}{Sprint\mbox{-}\nobreak\hspace{0pt}x\xspace}
\newcommand{\dash}{DASH\xspace}
\newcommand{\bdpf}{BDP_f}
\newcommand{\cwnd}{\text{cwnd}\xspace}
\newcommand{\bdp}{\text{BDP}\xspace}
\newcommand{\rtt}{\text{RTT}\xspace}
\newcommand{\obs}{inadequate window syndrome\xspace}

\usepackage[first=10, last=255]{lcg}

\newfont{\titlefont}{phvb8t at 17pt}
\newfont{\authorfont}{phvr8t at 12pt}
\newfont{\affilfont}{phvr8t at 11pt}
\usepackage[absolute]{textpos}
\newcommand{\ToAppear}{%
\begin{textblock*}{\textwidth}(0.95in,0.4in)
\begin{flushright}
    \noindent{\fbox{\textsf{Draft version---please do not redistribute.}}}
\end{flushright}
\end{textblock*}
}

\floatstyle{ruled}
\newfloat{program}{thp}{lop}
\floatname{program}{Program}

\usepackage[font=bf,aboveskip=5pt]{caption} % SPACE
\DeclareCaptionFont{mybf}{\bfseries}   % make math symbols bold
\renewcommand{\paragraph}[1]{\smallskip\noindent\emph{#1}}

\usepackage[hyphens]{url}
\ifthenelse{\equal{\COLOR}{yes}}{%
  \usepackage[colorlinks]{hyperref}%         % for online version
  \definecolor{darkblue}{RGB}{0,0,120}
  \definecolor{reallydarkblue}{RGB}{0,0,63}
  \definecolor{Black}{RGB}{0,0,0}
  \hypersetup{
    colorlinks,
    citecolor=Black,
    linkcolor=reallydarkblue,
    urlcolor=darkblue}
}{%
}
\usepackage{url}

\newcommand{\cut}[1]{}
\newcommand{\forcered}[1]{#1}
\ifthenelse{\equal{\COMMENTS}{yes}}
{%
\newcommand{\comment}[1]{\textcolor{red}{(#1)}}
\newcommand{\hil}[1]{\textcolor{red}{#1}}
\newcommand{\mjf}[1]{\textcolor{red}{(MJF: #1)}}
\newcommand{\sid}[1]{\textcolor{green}{(Sid: #1)}}
\newcommand{\ma}[1]{\textcolor{magenta}{(Mat: #1)}}
\newcommand{\note}[1]{\tnote{\ }{#1}}
\newcommand{\tnote}[2]{%
  \definecolor{randomcolor}{RGB}{255,100,100}
  \textcolor{randomcolor}{#1}\marginpar{\vspace{-0pt}\small\itshape\color{randomcolor}#2}
  }
}
{
\newcommand{\mjf}[1]{}
\newcommand{\sid}[1]{}
\newcommand{\ma}[1]{}
\newcommand{\comment}[1]{}
\newcommand{\hil}[1]{}

\newcommand{\note}[1]{}
\newcommand{\tnote}[2]{#1}
}

\newcommand{\eg}{{e.g.},\xspace}
\newcommand{\ie}{{i.e.},\xspace}

\newtheorem{myobs}{Observation}

\usepackage{enumitem}
\setlist{itemsep=0pt,parsep=0pt,leftmargin=1em}   % more compact lists
\usepackage{fancyhdr} 

\ifthenelse{\equal{\PAGENUMBERS}{yes}}{%
  \pagestyle{plain}
}{%
  \pagestyle{empty}
}

\usepackage[numbers,sort]{natbib}

\title{\vspace{2ex} \titlefont{} \TITLE \vspace{1ex}}
\author{
\authorfont{}{Matvey Arye$^*$, Siddhartha Sen$^{\dagger}$, Michael J. Freedman$^*$}\\ [4pt]
  \affilfont{$^*$Princeton University, $^{\dagger}$Microsoft Research} \\ [8pt]
}

\begin{document}
\maketitle

\ifthenelse{\equal{\SHOWTOAPPEAR}{yes}}{\ToAppear}{}
{\abstract% 
HTTP-based video streaming is a key application on the Internet today,
comprising the majority of Internet traffic today. Yet customers remain dissatisfied
with video quality, resulting in lost revenue for content providers. Recent
studies have blamed this on the adaptive bitrate selection (ABR)
algorithm used by client players, claiming it interacts poorly with TCP
when the video buffer is full, which causes it to underestimate 
available network bandwidth.

We show that the root cause of the problem lies in the data plane,
and that even a perfect control plane (ABR) algorithm is not enough to guarantee
video flows their fair share of network bandwidth.  Namely, it is the sequential download of
(small) video segments that is at fault, as they disrupt the normal interaction
between TCP congestion control and router queue occupancy. We analytically derive the
throughput of a video flow as a function of download size and network
conditions, and use this to develop an adaptive algorithm for selecting the
download size. Combined with pipelining, our approach achieves near-optimal
throughput and fast bitrate adaptation, regardless of the control plane
algorithm. We implement our approach as a \dash video player called \system, and
evaluate it against state-of-the-art proposals from the literature as well as 
deployed players from Netflix, Youtube, Hulu, and Amazon. \system consistently
achieves above 90\% of its fair-share throughput, while the previous
state-of-the-art exhibits high variability (\eg from 31\% to close to fair
share depending on the network conditions).
Industry players often achieve below 50\% of their fair share.

}

\section{Introduction}

Video is the main source of Internet traffic,
comprising a whopping 78\% of total North American traffic~\cite{cisco-video}.
%and expected to grow to 84\% by 2018
Yet poor video quality remains a source of
dissatisfaction for customers and lost revenue for content providers: 
%a report
%by Conviva estimates that content brands missed out on \$2.16 billon in revenue
%in 2012 for this reason alone~\cite{conviva}.
%a report by Conviva estimates that 
estimates place 58.4\% of views as impacted by low resolution in 2014~\cite{conviva2015}.

%- Merge next three paragraphs to talk about how it works roughly

Video streaming over HTTP is the dominant form of video consumption: it is
easy to deploy and allows content providers to reuse existing infrastructure
for
% MJF Include Hulu? MA: No, Hulu is not HTTP-based
content distribution.  Netflix, Youtube, and Amazon Video all use this form of 
streaming, and alone account for more than 50\% of all peak downstream
North American 
Internet traffic~\cite{sandvine}. There are several standards 
for HTTP-based  video
streaming, including proprietary ones from Apple, Microsoft, and Adobe, and the
open-source Dynamic Adaptive Streaming over HTTP (\dash) standard.  All use the
same underlying technique.  Movies are divided into segments of a given duration
(\eg 4 seconds of video),  and each segment is encoded at multiple pre-set
bitrates: higher bitrates result in larger, higher-quality segments. 
These segments are served as static content from regular web servers and caches. The video
player on the client determines when to download the next segment and at what
bitrate. 

We can divide a video player's functionality into a control plane and a data plane.
The control plane chooses when to download the next segment and uses adaptive
bitrate selection (ABR) to choose the segment's bitrate; it maintains the
downloaded segments in a buffer. The data plane downloads each segment via an
HTTP request.
% (it sits above HTTP/TCP).
%To make these decisions, the video player maintains a buffer of downloaded
%segments and uses adaptive bitrate selection (ABR) to determine the bitrate of
%the next segment.  p
Typically, the ABR algorithm selects a bitrate based on current buffer levels
combined with bandwidth estimates based on timing data from the previous downloads.  As long as
the video buffer is below the target level, segments are downloaded
sequentially (one at a time). When the buffer fills, downloads are paused until
the buffer drains below a certain watermark.  The ABR algorithm walks a
tightrope: if a selected bitrate is too high, the download may not
keep up with video playback, resulting in interruptions; if the bitrate is too
low, video quality and user satisfaction suffer.  Ultimately, the goal is to
pick a bitrate that matches the available network bandwidth.

Prior work has shown that video flows are unable to achieve their fair
share of available bandwidth when competing against other flows~\cite{festive,
Akhshabi12, Houdaille12, Huang12}. This is a common scenario: shared downlinks
are characteristic of both residential Internet
connections~\cite{sundaresan2011broadband, fcc:averages} and mobile
% MJF This is great data. That said, our eval should show competing video
% transfers, in addition to video vs. bulk? MA: Yes Section 6.4 has this (table 4)
networks~\cite{Jiang2012}. In 2014, households had an
\emph{average} of 1.5 devices 
streaming video concurrently during prime-time~\cite{conviva2015}. This
number is up 28\% since 2012, and other flows complete for bandwidth as well.

In a recent study, Huang et al~\cite{Huang12} attributed the problem to two
things:
(i) when the video buffer is full, the pauses between segment downloads cause TCP to
time out and reset the congestion window (\cwnd), and (ii) lower \cwnd values
cause the ABR algorithm to underestimate the available bandwidth, leading
it to select lower bitrates (smaller segments) that further stymie the growth of
\cwnd, creating a negative feedback loop.  Not surprisingly, the proposed
solutions have included alternative ABR algorithms---\eg Huang
et al's algorithm~\cite{Huang14} avoids filling the
video buffer---and techniques that ensure a minimum download size to
allow \cwnd to grow~\cite{Huang12}.

In common network conditions, however, the proposed solutions are
insufficient, given the manner in which network buffer sizes effect video
streaming performance. We develop a conceptual and analytical model of how the
iterative nature of HTTP-based video streaming interacts with TCP and network
buffering, and use it to devise a comprehensive solution.

%We argue, however,
%that this work did not identify the root cause of the
%problem, and hence the proposed solutions, while a clear improvement, are 
%incomplete or suboptimal in some way. Our goal is to thoroughly understand the
%underlying problem and use this insight to devise a comprehensive solution.
%
%\sid{Also could add reference to Section 4.1 and 4.2.}

We first review some TCP basics and establish a connection between a flow's
\cwnd and the bandwidth-delay product (\bdp) (\S\ref{sec:model}). 
%\sid{We observe that either \cwnd has to exceed BDP, or the flow relies on
%queued packets at routers to maintain a fair share of throughput.}  
Since competing flows increase the perceived roundtrip time by filling router
queues with their packets, they increase \bdp.  This combined with the sequential nature of
video segment downloads---which repeatedly drains the router queues of a
flow's packets---is what leads to suboptimal throughput
(\S\ref{sec:network}).  In particular, this is a data plane problem that
occurs even when the video buffer is not full,
%(the common case~\cite{akamaiSOI})
so it affects all control-plane algorithms, including those of Huang et
al~\cite{Huang12,Huang14}.  By addressing the problem, we can improve the
performance of all control planes simultaneously.

Armed with this insight, we devise a data plane solution based on
ensuring a minimum download size of video data (\S\ref{sec:solution}). 
Unlike prior solutions that take this approach, we use our knowledge of the
root cause to analytically derive the minimum download size required, as a
function of current network conditions, in order to achieve a $1 - \epsilon$
fraction of the video flow's fair-share throughput. 
%Deriving the minimim size dynamically is key to ensuring the ABR algorithms
%adaptability.

We describe two implementations of our solution: one uses expanded range
requests and runs inside a regular web page (\systemx); the other uses pipelined
requests and runs as a browser extension (\system).
% due to limitations in current browser APIs. 
%, which we also address (Section~\ref{sec:browser}).
%Current browser APIs inhibit pipelining from a regular web page, forcing us to
%implement \system as a browser extension; we suggest a possible addition to the
%browser API to address this (Section~\ref{sec:browser}).  
Both solutions are very simple on the surface, but right-sizing the
downloads is critical to their efficiency, which in turn relies on a correct
understanding of the problem. In particular, we show in \S\ref{sec:pipeline}
that simply turning on pipelining is not good enough.
%\system only needs to enforce the minimum download size when the
%video buffer is full. 
Our evaluation (\S\ref{sec:eval}) shows that our
solutions achieve large gains in throughput across a variety of control
plane algorithms. 
%uses the recent algorithm of Huang et al.~\cite{Huang14} mentioned above that
%was designed to solve the same problem, and we still show a significant
%improvement in achieved throughput. 

In effect, \system allows the control plane to focus on high-level
objectives such as quality of experience (QoE), while trusting the data
plane to execute its decisions efficiently. Our evaluation additionally 
shows gains in QoE metrics such as video bitrate, video stalls, and
subsequent rebuffering.
\ignore{
Since our solution only affects download sizes, it resides in the ``data
plane'' of video streaming, and can thus work with any ``control
plane'' (ABR) algorithm.  In fact, our evaluation (Section~\ref{sec:eval})
uses the recent algorithm of Huang et al.~\cite{Huang14} mentioned above that
was designed to solve the same problem, and we still show a
significant improvement in achieved throughput.
}

To summarize, we make the following contributions:

\begin{itemize}
\item 
  We show that sequential downloads are a first-class problem for
  video streaming performance, because they disrupt the normal interaction
  between TCP congestion control and router queue occupancy.

\item We develop a model to explain video flow throughput as a function of
download size and network conditions. We use this to define an algorithm that
adaptively determines the download size needed to achieve fair-share bandwidth
given estimates of current network throughput and RTT.

\item We implement our data plane solution, \system, as a \dash video player and
evaluate it in emulated and real environments against state-of-the-art proposals
and commercial players. We demonstrate \system's universality by applying it to
several control-plane algorithms.
%We also suggest an extension to the browser API that enables efficient video
% streaming within a web page.
\end{itemize}

\section{Background}

We first summarize prior efforts to explain the underperformance of video
flows. Then, we describe an analytical framework for diagnosing problems
in the data plane.

\subsection{Related work}
\label{sec:related}

Finding the performance of existing video players lacking, 
previous work largely proposed new ABR (control plane) solutions to improve
performance and QoE. These solutions are complementary to our data-plane
changes. Broadly speaking, these works considered three scenarios:
%can be divided into three categories based on the type of competing flows:

(i) When there are no competing network flows, previous work found that
video players often fail to optimize quality of experience (QoE) metrics such
as average video bitrate, rebuffering time, and frequency of bitrate
changes~\cite{Akhshabi11,Mok12,DeCicco10,Rao11}. Many ABR algorithms have been
proposed to optimize QoE metrics. For brevity, we note only two recent
algorithms that use a control-theoretic approach to optimize QoE~\cite{tian,
mpcdashabr}. These ABR algorithms were not designed (or evaluated) for cases
when video flows compete for bandwidth, and thus do not address the problems
caused by client-side pauses.  

(ii) When multiple video players compete for bandwidth, 
client-side pauses that occur when the video buffers become full can become
synchronized~\cite{festive, Akhshabi12, Houdaille12, probeandadapt}.  Such
synchronization can make it hard for the video players to accurately sense
available bandwidth and trap one player in a lower-bandwidth setting, leading
to unfairness. FESTIVE~\cite{festive} and PANDA~\cite{probeandadapt} address
these problems by injecting randomness into pause scheduling to
de-synchronize video flows, and
propose new methods of pause-aware bandwidth estimation.

(iii) When a video player competes against a bulk flow, client-side pauses can
lead to TCP congestion window timeouts and other effects that favor the bulk 
flows~\cite{Huang12}.  Huang et al.~\cite{Huang14} proposed
an ABR algorithm that will not fill the buffer and incur
client-side pauses unless the player is already downloading segments
at the highest bitrate available. This minimizes the number of such pauses
that can occur. 

Yet video players can fail to achieve
their network fair-share even when using the perfect ABR algorithm
(\S\ref{sec:network}), \ie one
that picks a bitrate to exactly match the correct fair-share and eliminates all
client-side pauses.
Thus, changing the control plane while using an unmodified
data plane is insufficient.  Our work is a data-plane solution that changes
different operational parameters than previous ABR solutions. In
particular, we change the size of data transferred per request 
or per pipeline ``train'' (if using pipelining), and show that
%Unlike in current systems, we define this size
%independently of the video bitrate and determine it dynamically based on
%current network conditions.  
this achieves network fair-share with any control plane. 

Operating an ABR algorithm on top of an unmodified data plane is complicated
by the fact that the ABR decision affects network throughput: smaller bitrates
(segment sizes) and/or longer pauses cause throughput to drop. This
interaction leads to a negative feedback loop when lowering video
bitrate~\cite{festive, Huang12} and contributes to the difficulty in bandwidth
estimation (a problem noted in~\cite{Huang12, probeandadapt, festive}).  Our
data plane solution breaks this interaction by guaranteeing fair-share of
network throughput independent of the ABR algorithm. 
%We leave open the question
%whether some ABR algorithms can exhibit the synchronization problem between two
%video players described in (ii), although we did not see this effect during our
%experiments.

Previous work has proposed less-thorough versions of our solution based on
empirical observation, \eg Huang et al.~\cite{Huang12} advocate downloading larger
(fixed-sized) chunks of video data. Yet as we show, the correct size 
depends on network conditions, and no single size is appropriate for all
scenarios.  
%Picking a size that is too small will negatively effect throughput
%while large sizes prevent the player from adapting video bitrate quickly.
Another study~\cite{dashhttp2} noted that HTTP2, which uses
persistent connections and pipelining, improves video performance. While 
simple pipelining may lead to some benefit, it doesn't guarantee network
fair-share.  In contrast, we derive the number of requests to pipeline together
analytically as a function of network conditions so as to guarantee network fair-share.
%We also specify how to schedule the requests
%within a pipeline so that ABR algorithm can adapt the bitrate quickly without
%compromising throughput.
%We consider only client-side solutions 

MSPlayer~\cite{msplayer} proposes dynamic chunk sizes
to synchronize downloads across parallel video streams, not to increase
throughput. Other work improves video streaming performance by modifying on-path
network elements, \eg~\cite{Houdaille12, trickle}.
We focus on client-side solutions as those are easier to deploy. 

\subsection {TCP Throughput Basics}
\label{sec:model}

\begin{figure}[t]
 \centering
  \includegraphics[width=0.6\linewidth]{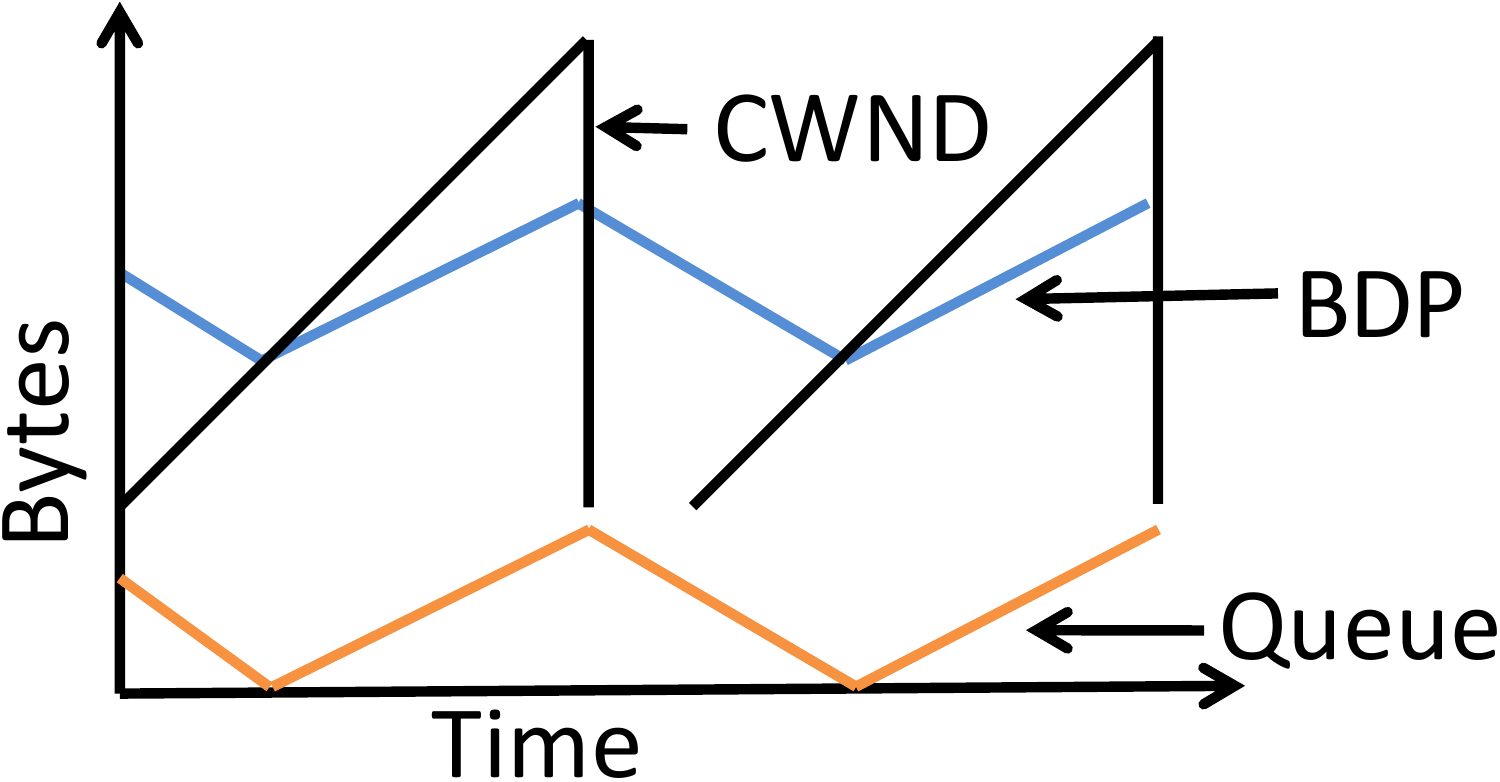} 
 \caption{A simplified schematic of TCP dynamics.}
\label{fig:tcp}
\end{figure}

To understand what goes wrong in the data plane of video flows, we need
to review some TCP basics.  Recall that TCP limits the number of unacknowledged bytes sent on the
network to the congestion window (\cwnd). Since an acknowledgment takes 
a roundtrip time (\rtt) to arrive, this limits the flow's throughput to $cwnd /
\rtt$.  Equivalently, to support a throughput of $T$, \cwnd must exceed
$T \times \rtt$ bytes, the familiar bandwidth-delay product (\bdp).  When
multiple flows share a bottleneck link, the \cwnd of each flow has to exceed the
\emph{fair-\bdp}---the fair share of the link bandwidth multiplied by \rtt.
(For simplicity, we use BDP and fair-BDP interchangeably.)  It is critical to
note that \rtt is a dynamic quantity because it includes queuing delays in the
network. Consequently, \bdp is also
dynamic: it rises and falls as network queues fill and drain (see
Figure~\ref{fig:tcp}).  These fluctuations can be large: a 3Mbps link with a
256KB queue (representative of U.S. home connections, see
\S\ref{sec:network-params}) can take 683ms to drain in the worst case.
If \rtt without queuing is 100ms, BDP could grow by 583\%.

TCP flows do not always rely on \cwnd exceeding \bdp in order to get their fair share of
throughput. Instead, in-network queues absorb the peaks and troughs
of TCP's famous \cwnd sawtooth pattern.
%\cite{tcpQ, Guido04}. 
Figure~\ref{fig:tcp} shows the
interaction between \cwnd and in-network queueing. When \cwnd exceeds \bdp, the
sender is transmitting more bytes than the bottleneck link can drain, so network
queues start growing.  Eventually, the queues fill to capacity, packets are 
dropped, and TCP's congestion avoidance cuts \cwnd in half.  Now, 
\cwnd might be less than \bdp, but the full bottleneck queue
can supply data, so no throughput is lost.  If the queue is sized
appropriately, it will finish draining just as \cwnd again exceeds
\bdp.
%~\cite{Guido04}. 
This analysis of the interaction between congestion
control and in-network queuing applies to all AIMD variants of 
TCP congestion control including Cubic,
Reno, and Tahoe. 

The following observations, while not new, are 
critical to understanding the problems with today's video flows.
\begin{myobs}
BDP is a dynamic quantity that rises and falls as network queues
fill and drain.
\label{obs:bdp}
\end{myobs}

\begin{myobs}
To fully utilize network bandwidth, either \cwnd $>$ \bdp or network queues must 
  not be empty. We call a violation of this the {\bf \obs}.
\label{obs:fs}
\end{myobs}

As we will see, video flows tend to exhibit the \obs,
leading to poor throughput.

\section{Explaining Poor Video Performance}
\label{sec:network}

We can now (re-)explain video flow performance and
recharacterize prior conclusions. There are three
scenarios of interest for a video flow: (i) no competing flows, (ii) competing
against a bulk flow with pauses between requests, and (iii)
competing against a bulk flow without pauses. \forcered{(Our evaluation also
considers non-bulk flows.)}
%MAT: Consider adding ``Our evaluation also considers non-bulk f

Prior work has focused on the first two scenarios.
Huang et al.~\cite{Huang12} showed that video flows underestimate
their fair share of bandwidth when competing against bulk flows, causing the
video player to choose a lower bitrate than necessary. They attributed this
to the periodic pauses that occur when the video buffer is
full: the player alternates between downloading video segments to fill the
buffer, and pausing while the buffer drains due to video playback.
During a pause, \cwnd times out and collapses to its initial value (\eg 10
packets, the Linux default).
%\footnote{The default in many operating
%systems, including Linux~\cite{rfc:3390}.}). 
Since the bulk flow keeps filling the network queues, the video flow
experiences high packet loss upon resuming download, causing \cwnd to get
``repeatedly beaten down.'' 
They did not observe this effect when there was no competing flow.

Our analytical model explains these effects.
Moreover, we discover the
problem occurs \emph{even when there are no pauses in
downloading} (scenario (iii) below).  This occurs while the video
buffer is filling and in general when the fair share of bandwidth is below the chosen video bitrate.
According to Akamai~\cite{akamaiSOI},
%'s State of the Internet Report for Q4 2014
this is common for high quality video: only 18\%
of network flows in the U.S. have an average speed above the 15Mbps bitrate characteristic of
4K video. 
\ma{Make unit spacing e.g. 15Mbps vs 15~Mbps consistent throughout.}
We find that in this scenario the video flow also fails to achieve its
fair share of network throughput even when using the solutions proposed in
previous work.

We now (re-)explain the three scenarios:

\textbf{(i) No competing flows.} Without any competing flows, there is
no queueing delay in the network, so \bdp remains lower than \cwnd. This
satisfies Observation~\ref{obs:fs}.  Even if a pause occurs and \cwnd drops to
its initial value, this is still often higher than \bdp.  For example, in
Huang et al.'s~\cite{Huang12} experiment setup, the \bdp of a 5Mbps link with no
queueing is 100 kbits, while the initial \cwnd is 117 kbits. This explains
why they observed good performance when there were no competing flows.

\textbf{(ii) Competing against a bulk flow with pauses between requests.}
Competing flows induce a queueing delay, and thus raise BDP during a
pause in the video flow.  As we observed earlier, this increase can be dramatic.
When the video flow resumes downloading, its initial \cwnd of 10 packets falls below the current BDP; meanwhile, the preceding pause has drained
all video packets from the bottleneck queue. Together, these conditions violate
Observation~\ref{obs:fs} and thus exhibit the \obs, resulting in suboptimal
throughput.  This explanation is the same as Huang et al.'s~\cite{Huang12}.
%As we show in Section XXX, this analysis allows us to use the value of the BDP
% to quantify this effect.

% SID: Removed this figure, as it is just a summary of prior work which we can
% write in text more succinctly.

\ignore{
\begin{figure}[t]
 \centering
  \includegraphics[width=\linewidth]{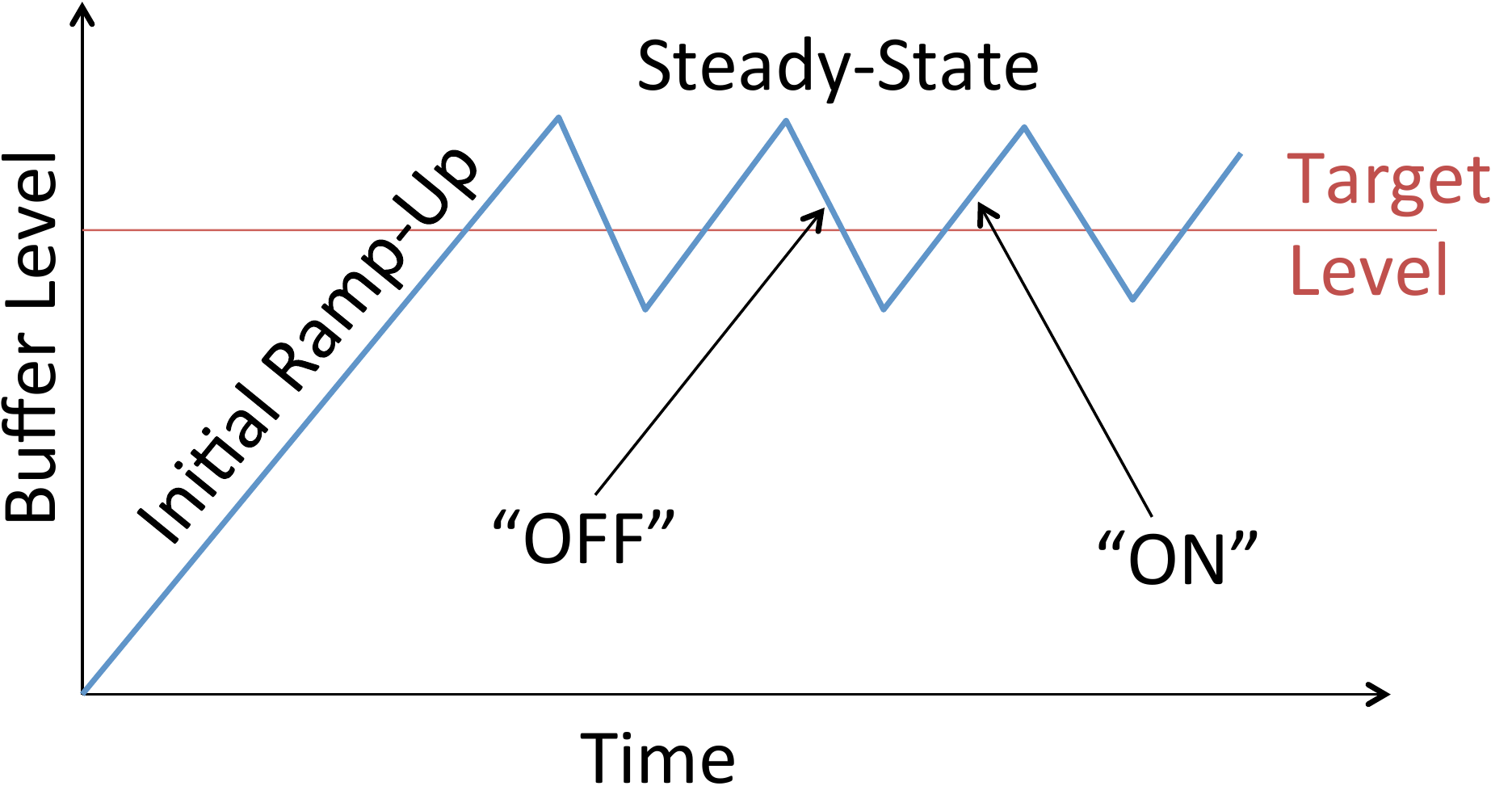} 
 \caption{A schematic of the video buffer level over time.}
\label{fig:buffer-level}
\end{figure}
}

\begin{figure}[t]
 \centering
  \includegraphics[width=0.60\linewidth]{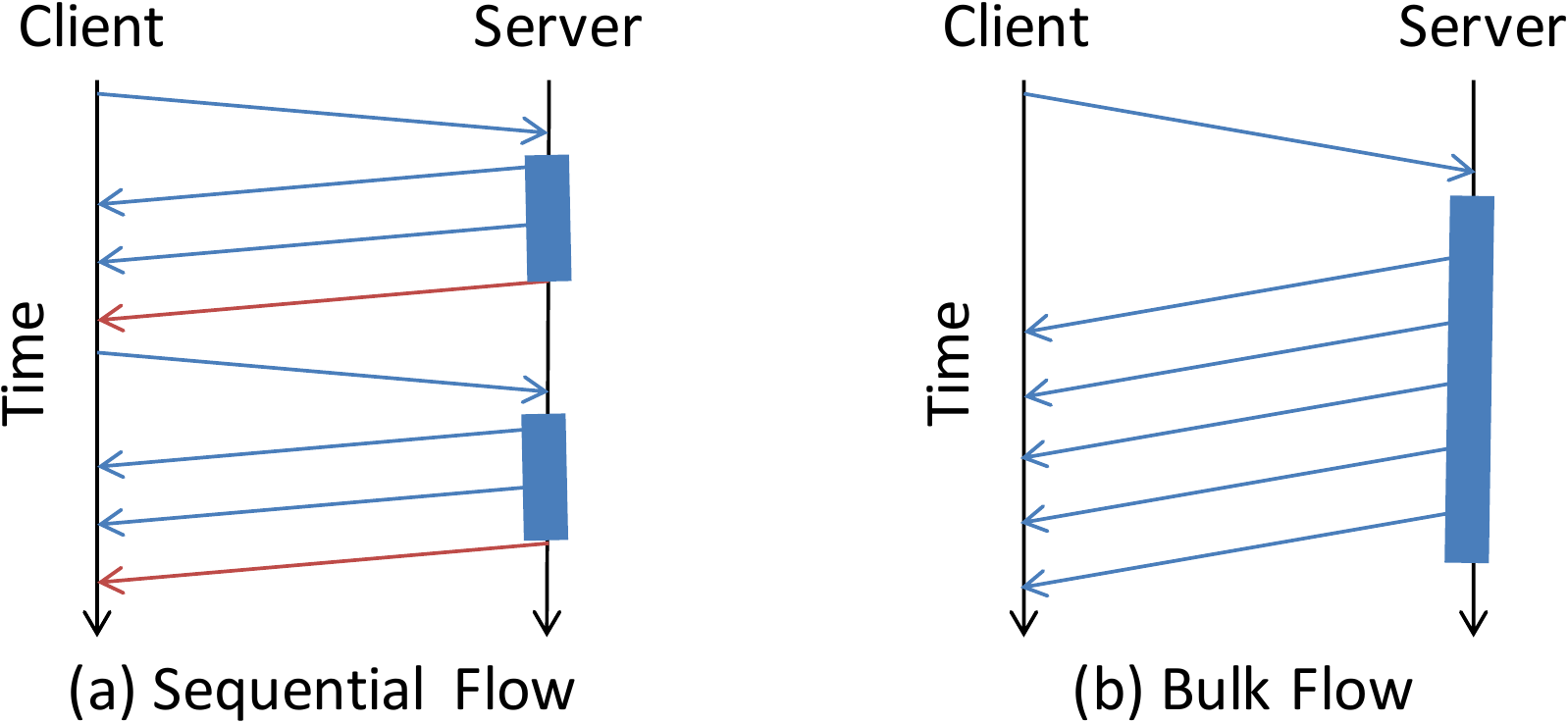} 
 \caption{A schematic of sequential vs.\ bulk flows.}
\label{fig:bulk_iterative}
\end{figure}

\textbf{(iii) Competing against a bulk flow without pauses between requests.}
Even without pauses, the video player still periodically drains the network
queues of all video packets. This is because it downloads the video segments
sequentially using HTTP requests.  Figure~\ref{fig:bulk_iterative} illustrates
the difference between this sequential flow and a bulk flow.  The video player
waits to receive the last packet of the previous request (shown in red) before
issuing the next request.  The act of receiving this packet {\em drains the
network queues of all video packets}, as the video server has no more data to
send until it receives the next request. If \cwnd is below \bdp at this
time---\eg right after it halves during congestion avoidance, as shown in 
Figure~\ref{fig:tcp}---then the flow will exhibit the \obs and 
achieve suboptimal throughput.

%This problem of small sequential downloads is inherent to the data plane and
%can thus affect any control plane algorithm. 
By focusing on control-plane issues and QoE metrics, prior work may have
overlooked the third scenario; also, some experiments used 
queue sizes that were too small, \eg 15KB in Huang et al.. A small
bottleneck queue may not allow BDP to increase enough to
cause the \obs.  The majority of U.S. homes have downlink 
queues greater than 128KB (\S\ref{sec:network-params}), however, and
bufferbloat~\cite{bufferbloat} remains a problem.

Prior work has found a negative feedback loop
when video flows achieve less than their fair
share of throughput~\cite{Huang12,festive}:
lower throughput causes the ABR algorithm to switch to a
lower bitrate; lower bitrate segments are smaller 
%(for the same number of video seconds),
so less data is downloaded and \cwnd grows less; lower \cwnd
values lead to lower throughput. Although previously observed for the second
scenario above, we find that it also holds for the third.

%We focus on throughput since the problem we have identified is a data plane
%problem.
	
\subsection{Empirical validation}

\begin{figure*}[t]
 \centering
 \begin{subfigure}[]{0.32\textwidth}
    \includegraphics[width=\linewidth]{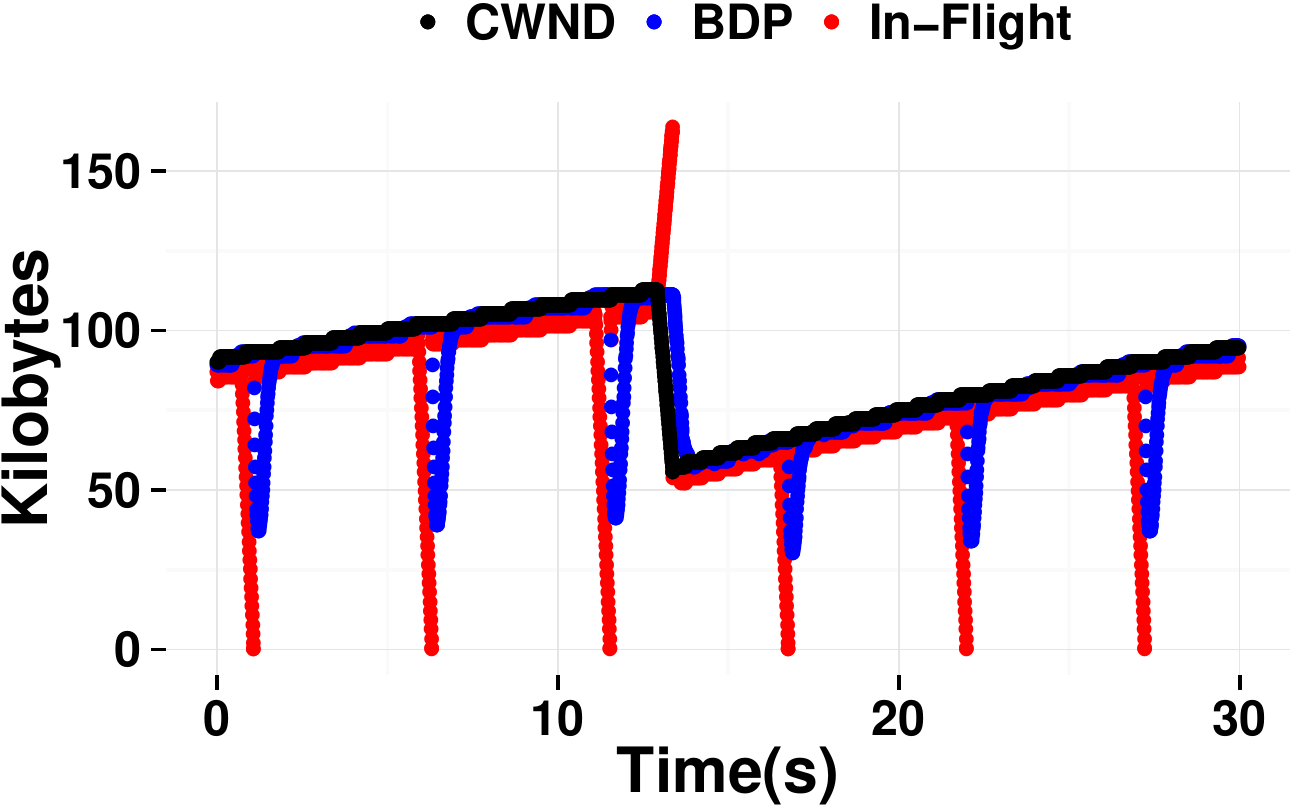} 
    \caption{No competing flows, no pauses}
    \label{fig:cwnd_noiperf}
 \end{subfigure}
 \begin{subfigure}[]{0.32\textwidth}
    \includegraphics[width=\linewidth]{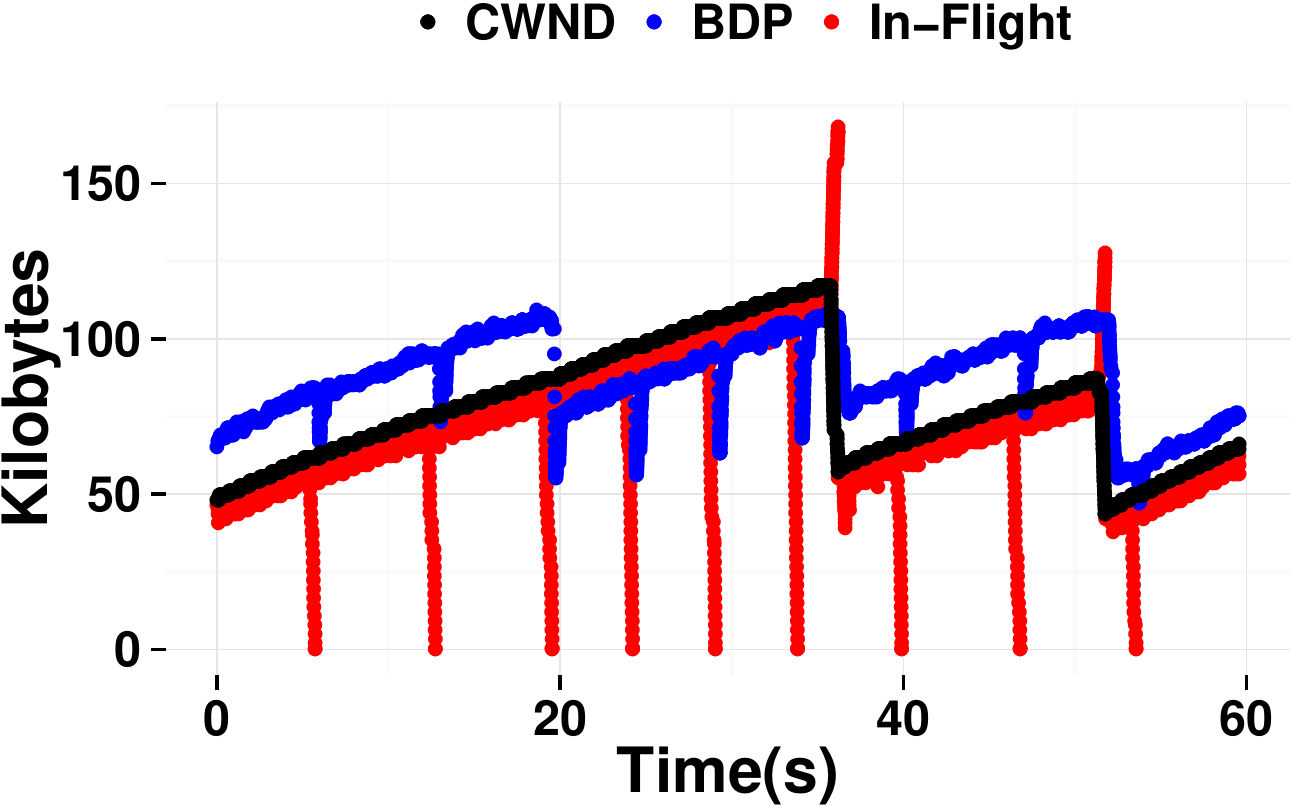}
    \caption{Competing with iperf, no pauses}
    \label{fig:cwnd_iperf}
 \end{subfigure}
 \begin{subfigure}[]{0.32\textwidth}
    \includegraphics[width=\linewidth]{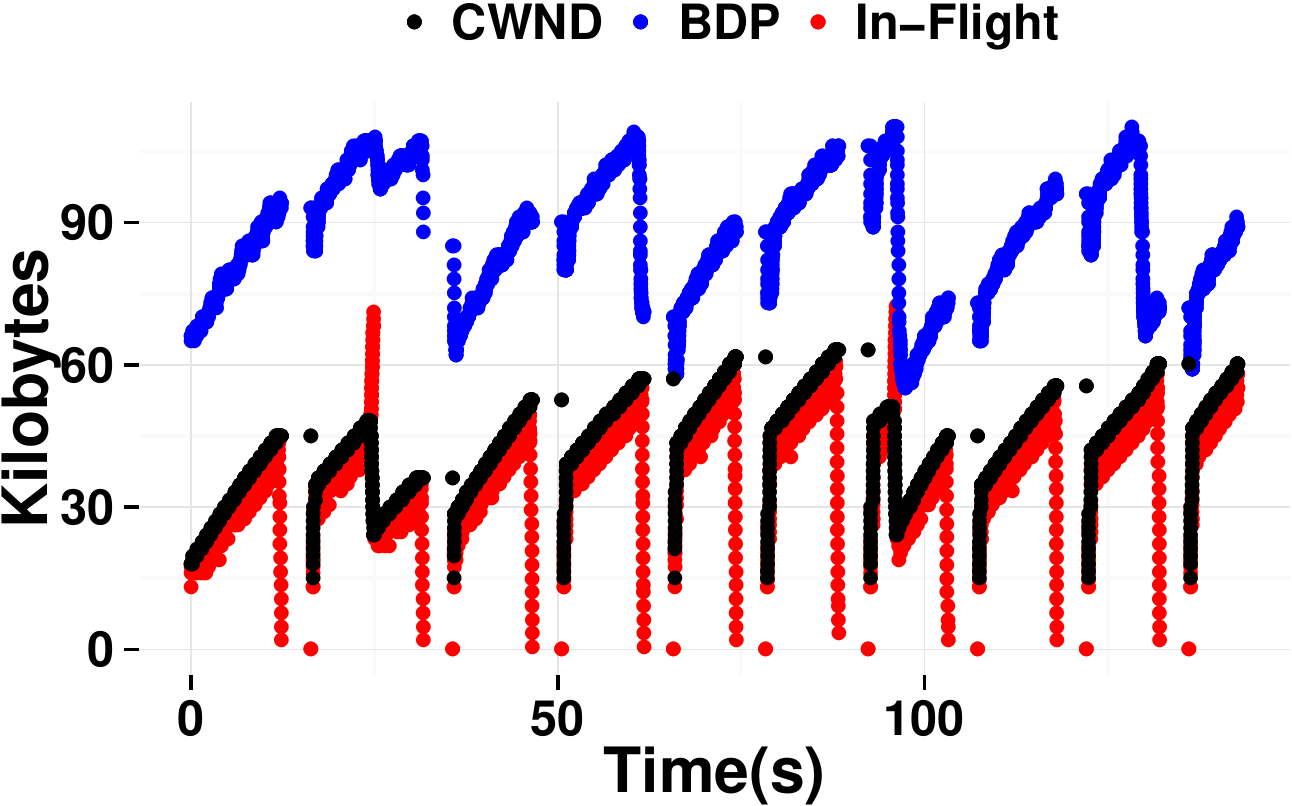}
    \caption{Competing with iperf, 4s pauses}
    \label{fig:cwnd_iperf_pause}
 \end{subfigure}
 \caption{Behavior of three video
 flows.  For each, the fair share of 
 bandwidth is 2Mbps, propagation delay is 20ms, router queue size
 is 100KB, and video segment size is 1250KB.
 Fair-BDP calculated using TCP's RTT estimate and the fair-share
bandwidth. 
 % queue size 400ms 
 %The bytes in-flight is the  difference between the last sequence number sent
 %and the most recent cumulative ACK received. When the amount of bytes
 % in-flight exceeds BDP, this indicates the beginning of loss and use of SACK.
 }
\label{fig:cwnd}
\end{figure*}

\ignore{
\begin{figure}[t]
 \centering
  \includegraphics[width=\linewidth]{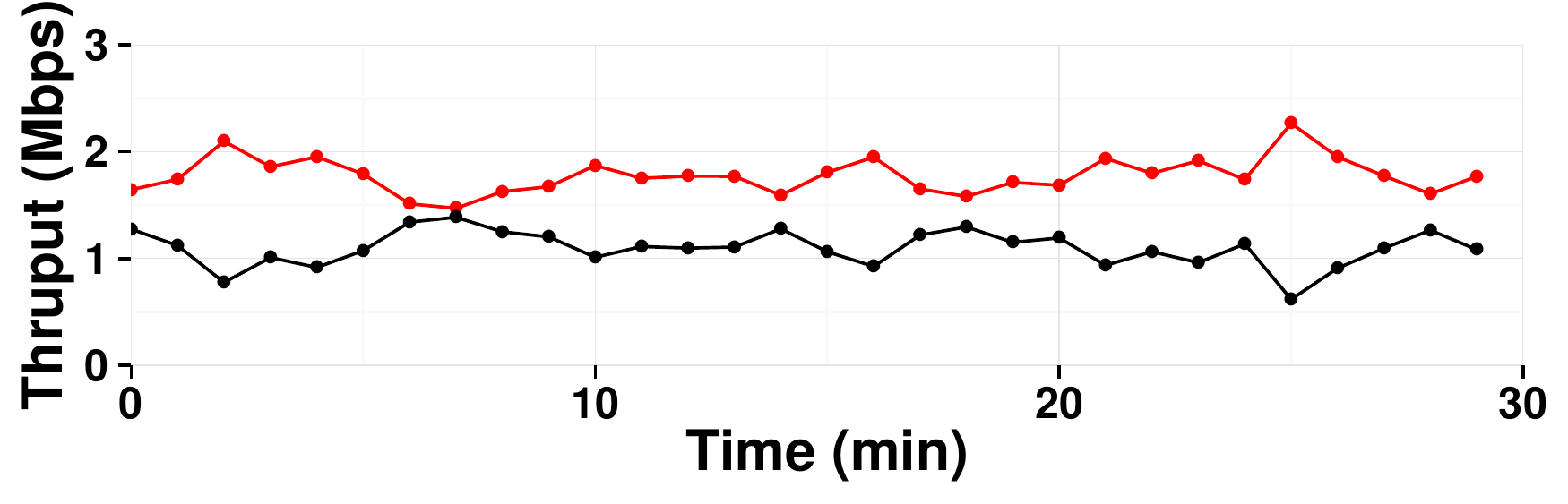} 
  \includegraphics[width=\linewidth]{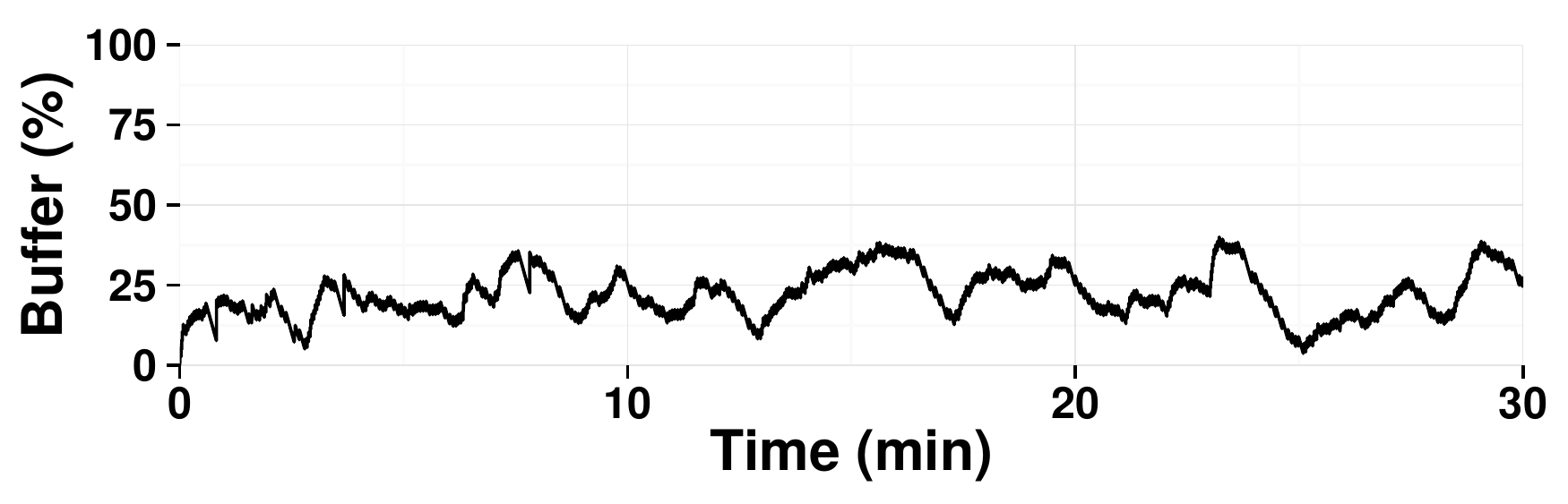} 
  \caption{Aggregate throughput and video buffer level for a
    video flow (black line) competing against a bulk flow (red line). The 
    bottleneck link bandwidth and queue size are 3Mbps and 256KB, respectively.}
\label{fig:dashperf}
\end{figure}
}

\begin{figure}[t]
 \centering
  \includegraphics[width=\linewidth]{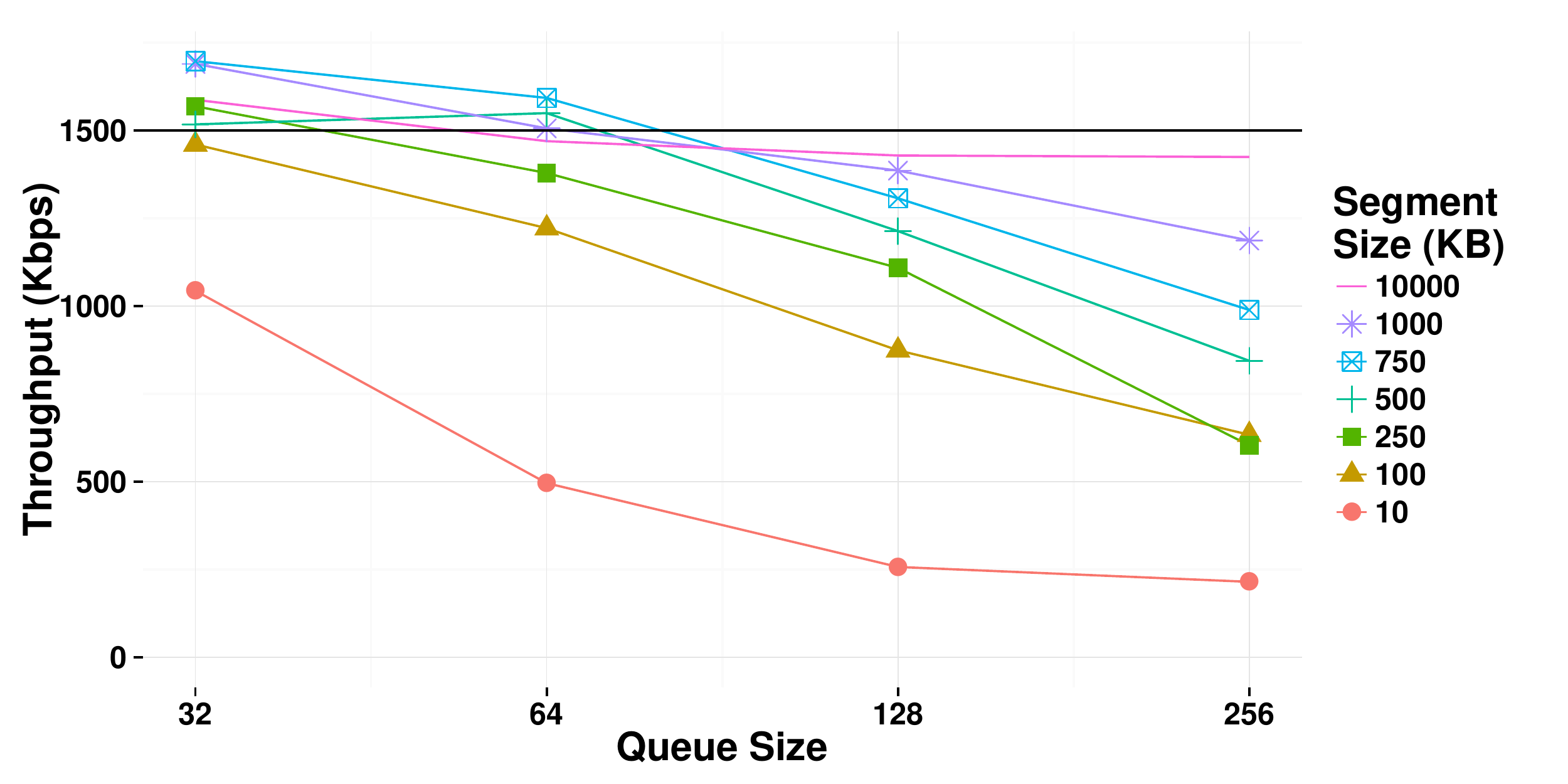} 
  \caption{ 
  Throughput of video flow using sequential downloads (without
  pauses) as segment size varies, while competing with an iperf flow.
  The fair share of bandwidth is 1500Kbps. 
  The throughput is taken as the median value of 200 transfers.
  }
\label{fig:fragment_size}
\end{figure}

Using \texttt{tcp\_probe} traces of actual network transfers, we can validate our explanations of
video performance. 
Figure~\ref{fig:cwnd} traces a video flow in each of the above 
scenarios.
%We used the tcp\_probe Linux kernel module to obtain internal TCP information.
%during the 
%transfers in our lab's controlled environment.

Without competing flows (Figure~\ref{fig:cwnd_noiperf}), the video
flows encounters no problems because \cwnd remains above \bdp.  The
situation is starkly different when competing against a bulk flow with pauses
(Figure~\ref{fig:cwnd_iperf_pause}). Here \cwnd times out repeatedly due to the
pauses and never reaches fair-\bdp. The resulting \obs leads to poor throughput.

Figure~\ref{fig:cwnd_iperf} shows the scenario that has been overlooked,
where there is a competing flow but no pauses in downloading. 
We still see the \obs:
\cwnd falls below fair-BDP at the same time the router queues are emptied of
all video packets. This happens at the boundary of video segment downloads: the
end of the previous download drains the router queues of all video data
(in-flight packets drops to zero), while the queueing delay induced by the
competing flow prevents fair-\bdp from falling as much as it did without a
competing flow.  Together, these  factors make it likely that, at the start
of a next segment download, \cwnd is below fair-\bdp. This is exactly what
happens at seconds 5, 11, 40, 48 and 53.  Thus even without pauses, the
sequential nature of video downloads can lead to suboptimal throughput.
Our evaluation shows that this degradation can be severe in practice, especially
for industry video players (Figure~\ref{fig:compare_app}). 
\ignore{
Figure~\ref{fig:dashperf} shows a representative example of such a scenario from
our evaluation. In this case, the target video buffer level (120 seconds) is
never reached, so no pauses are occurring, and yet the video flow achieves only
79\% of its fair share on average compared to the bulk flow.
}

Finally, we validate our claim that the negative feedback loop previously
observed applies even without pauses.  Figure~\ref{fig:fragment_size} shows the
throughput of sequential downloads as segment size (bitrate) decreases.
Clearly, as the ABR algorithm selects lower bitrates, performance will
continue to spiral downwards.  

\subsection{Towards a solution}

Previous work proposed ABR algorithms that address the second scenario by
eliminating pauses between requests. For example, Huang et al.'s
algorithm~\cite{Huang14} chooses the bitrate based on the current buffer level,
and thus avoids filling it unless the available bandwidth supports 
the maximum bitrate. 

However, eliminating pauses is not sufficient: the problem of sequential 
downloads exists even without pauses, and therefore cannot be resolved by
changes in the control plane ABR algorithm. Instead, we must change the way
segments are downloaded in the data plane. 
%The next section develops our data plane solution, which we call \system.

%This prevents the negative feedback
%loop that occurs with standard ABR algorithms and thus improves performance.

Our data-plane solution, \system, insulates ABR algorithms from feedback
interactions with the network layer. This allows ABR algorithms to focus on QoE
metrics instead of the negative feedback loop problems above.
Even standard ABR algorithms that had previously shown poor network throughput
performance are able to achieve their fair-share of throughput when running on
top of our data plane.

\section{Fixing Video Performance}
\label{sec:solution}

Armed with the above analysis, we introduce a new data plane that avoids the
interruptions caused by sequential downloads.  We achieve this by increasing the
amount of data that is downloaded as a continuous stream, which we call a
\emph{chunk}.  A chunk that spans multiple video segments allows \cwnd to grow and avoids draining the network
queues between segments, satisfying Observation~\ref{obs:fs}. In order
to use chunks effectively, we need to determine the minimum chunk size needed 
to achieve fair-share throughput (\S\ref{sec:adaptive}), while still allowing the
control plane to adapt the bitrate to available bandwidth
(\S\ref{sec:dataplane}).

\subsection{Adaptive chunk sizing}
\label{sec:adaptive}

To determine the appropriate size for a video chunk, we first quantify the
relationship between chunk size and network throughput.
%, drawing from our analysis in Section~\ref{sec:network}.
We then estimate the minimum chunk size needed to achieve a $1 -
\epsilon$ fraction of the fair-share throughput. 
\ignore{
Although prior work has advocated using larger chunks (\eg ~\cite{Huang12}),
they did not identify the root cause of the problem as we did in Section~\ref{sec:network},
and hence did not derive the chunk size analytically or adaptively.
}

%Finally, we present an algorithm that allows a video player to dynamically
% determine the right chunk size to use.  

%{\bf Relating chunk size to network throughput.}
We define the {\em efficiency} of a chunk transfer as the ratio between the
achieved and fair-share throughput.
%, given the
%current network conditions.
%On a high level, we derive this estimate as
%the ratio between the amount of bytes actually leaving the bottleneck link and
%what that amount would be if the flow had not underused the link by having \cwnd
%below BDP at the start of the segment request.  
%
% MJF this doesn't make sense anymore, as graph change. 
%Intuitively, the numerator (actual throughput) in
%the efficiency ratio follow the green line in Figure~\ref{fig:tcp} while the
%denominator (fair-share throughput) follows the orange line throughout the
% transfer.
We estimate achieved throughput by estimating the number of round-trips
used to transfer the chunk during the three phases of TCP:
%To estimate the achieved throughput, we estimate the number of round trips
%required to transfer the chunk. 
%There are three phases in the transfer:

\ignore{
We define the {\em efficiency} of a chunk transfer as the
ratio between the actual achieved network throughput and the network fair-share for a given chunk
size and network conditions.
%On a high level, we derive this estimate as
%the ratio between the amount of bytes actually leaving the bottleneck link and
%what that amount would be if the flow had not underused the link by having \cwnd
%below BDP at the start of the segment request.  
%
% MJF this doesn't make sense anymore, as graph change. 
Intuitively, the numerator (actual throughput) in
the efficiency ratio follow the green line in Figure~\ref{fig:tcp} while the
denominator (fair-share throughput) follows the orange line throughout the transfer. 

To estimate the actual throughput of the transfer, we first estimate
the number of round-trips it takes to complete the transfer of a chunk.
There are three phases in the transfer:
}

\begin{enumerate}[itemindent=0pt,leftmargin=1em]
  \item \textbf{Slow start.} From the beginning of the transfer to
  the slow start threshold (SST), the number of bytes transferred doubles every
  round trip. Given an initial \cwnd of 10 packets and the MSS, 
  $10\times MSS$ bytes are transferred in the round trip. Therefore the
  number of rounds in this phase is: 
    $ r_1 = \lceil \log_2(SST/(10\times MSS) \rceil + 1 $
    and via a geometric series the total bytes transferred is:
    $ b_1 =  (10\times MSS) (2^{r_1}-1) $.
  \item \textbf{Additive increase.} From the slow start threshold until 
    \cwnd reaches fair-BDP (denoted $\bdpf$),
    \cwnd increases by one MSS every RTT. Thus, the number of rounds is:
    $ r_2 = \lfloor (\bdpf-SST)/MSS \rfloor + 1 $
    and the total bytes transferred is:
    $ b_2 =  r_2 \times SST+MSS (r_2-1)$.
  \item \textbf{Transfer at BDP.} For the rest of the transfer, the 
  bytes transferred per RTT is equal to fair-BDP. Given
  chunk size $S$, the number of rounds is: 
    $ r_3 = \lceil (S-(b_1+b_2))/\bdpf \rceil $.
\end{enumerate}

\noindent The total number of rounds is $r = r_1+r_2+r_3$, and 
%(We make the appropriate correction if the size of the chunk is so small that the transfer is completed
%before phase 2 or phase 3.) 
the average throughput per RTT is $S/r$.
%, which can also be viewed as the average \cwnd.
% out of the bottleneck link. 
Since the fair-share throughput per RTT is simply fair-BDP, efficiency is:
%of the chunk transfer is:
\begin{equation}
 E = S / (r\times \bdpf)
 \label{eq:efficiency}
\end{equation}
This analysis shows that smaller chunks transfer at lower efficiency
because the fraction of time spent in the first two phases (before \cwnd
reaches fair-\bdp) is higher.  On the flipside, keeping chunk size constant, the
efficiency of the transfer decreases as fair-\bdp increases because it takes
longer for the \cwnd to reach it ($r_1$ and $r_2$ increase).

\ignore{This analysis shows that the throughput of a video flow depends on the
length of the segments and network conditions. Shorter segments will have a
lower efficiency ratio as the fraction of the time they spend in the first two
phases of the transfer (before \cwnd reaches fair-BDP) will be higher.  At the same
time, keeping segment sizes constant, the efficiency ratio will decrease
as fair-BDP increases because it will take longer for the \cwnd to reach it (increasing $r_1$
and $r_2$).
}

\begin{figure}[t]
 \centering
  \includegraphics[width=0.75\linewidth]{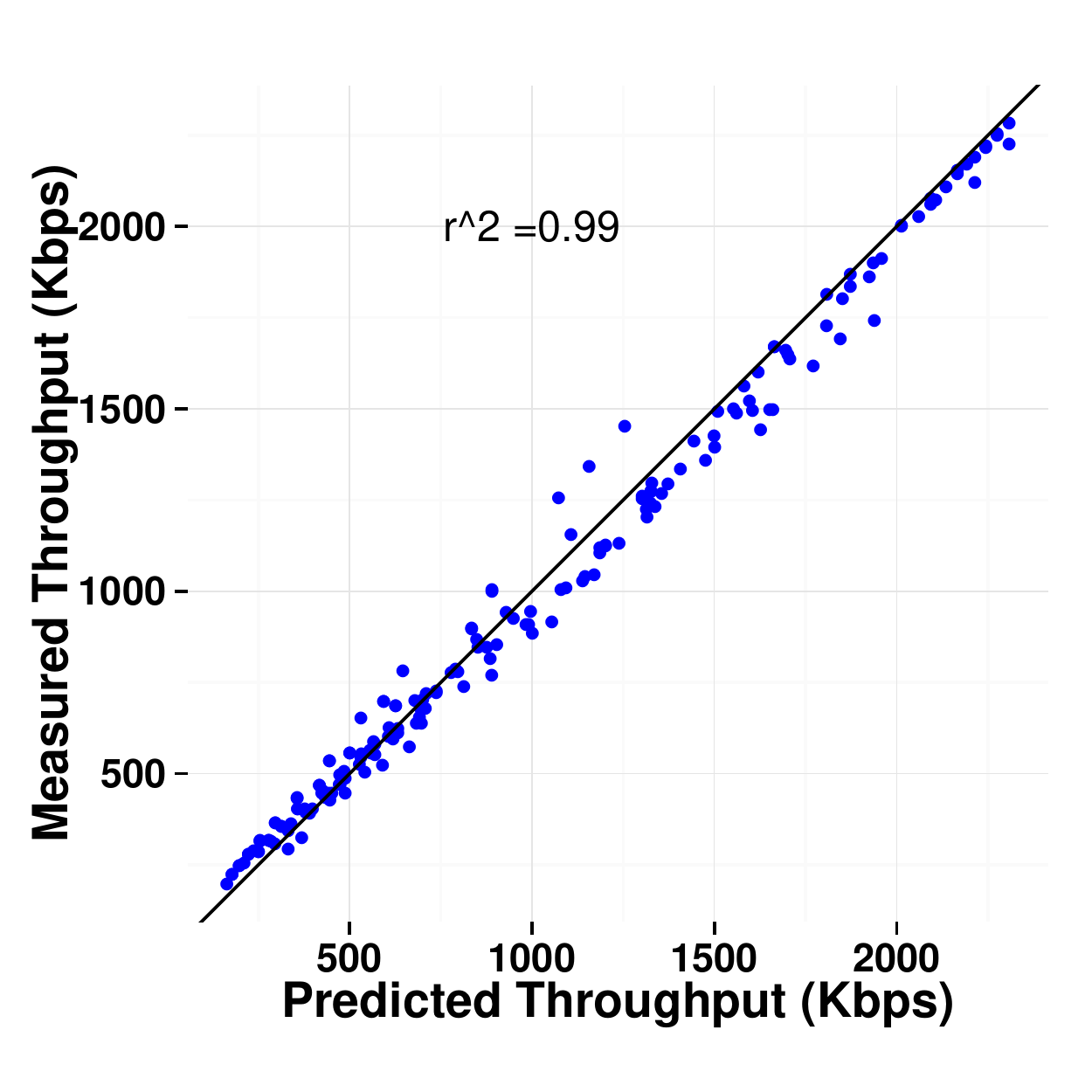} 
 \caption{
 Predicted vs. measured throughput for video chunk downloads over
 a bottleneck link with bandwidth 2.5Mbps, queue size
 either 250KB or 31KB, and propagation delay ranging from 100-1000ms.
 Chunk sizes ranged from 117-1800KB.}
% %800 ms and 100ms queue size
\label{fig:pred}
\end{figure}

To validate Equation~\ref{eq:efficiency}, we use it to predict the throughput
of real chunk transfers. Figure~\ref{fig:pred} shows that the prediction is
accurate across a wide range of chunk sizes and bottleneck queue sizes.  The
prediction takes as input the measured average slow-start threshold, chunk size,
and bottleneck link bandwidth. Each data point is an average of 10 chunk
transfers.
%

%Each experiment transfers 10 chunks
%with a 5 second pause between each transfer to allow TCP to return to
%slow start, and reports the average of the achieved throughputs.

Using Equation~\ref{eq:efficiency}, we determine the chunk size large enough
to ensure an efficiency of $1 - \epsilon$, for any $\epsilon$.  This
ensures the video flow achieves at least $1 - \epsilon$ fraction of its
fair-share throughput (in our experience, $\epsilon = 0.1$ is sufficient).
%\ignore{
%Specifically, chunk sizes need to be
%big enough so that that the efficiency ratio is close to $1$ (\ie $1-\alpha$). 
%This efficiency ratio aims to ensure that the flow receives at least
%($1-\alpha$)\% of its fair share of bandwidth.\footnote{In our
%experience an $\alpha = 0.1$ works well.}
%}
Equation~\ref{eq:efficiency} can be rewritten as:
$$E=\frac{S}{r \times \bdpf} = \frac{S}{(r_1+r_2+r_3) \bdpf}$$

If we bound $(r_1+r_2)/(r_1 + r_2  + r_3) \le \epsilon$ we get:
$$ 1 - \epsilon \le \frac{r_3}{r_1+r_2+r_3} = \frac {r_3
\bdpf}{(r_1+r_2+r_3)\bdpf} $$

\sid{Slack seems problematic} Since $r_3 \bdpf = S - (b_1+b_2)$, which is less
than $S$, $$ 1-\epsilon < \frac{S}{(r_1+r_2+r_3)\bdpf} = E $$

%Substituting for $r_3$ and simplyfing we get:
%$$ U= \frac{S}{S+(r_1 + r_2) \bdpf - (b_1+b_2)} $$
%
%If we bound $(r_1+r_2)/(r_1 + r_2  + r_3) \le \alpha$ we get the following inequality
%by substituting for $r_3$:
%$$ 1 - \alpha \le \frac{r_3}{r_1+r_2+r_3} = \frac {r_3 \bdpf}{(r_1+r_2)\bdpf + S-(b_1+b_2)} $$
%
%Since $r_3 \bdpf < S$, we get:
%$$ 1-\alpha < \frac{S}{S+(r_1 + r_2) \bdpf - (b_1+b_2)} = U $$

Thus, if we set $r = (r_1+r_2) / \epsilon$, then the efficiency has
to be greater than $1-\epsilon$. Intuitively, we bound the number of
round trips during which throughput underperforms ($r_1+r_2$) to a small
fraction ($\epsilon$) of the overall transfer ($r$).

\begin{program}[t]
\footnotesize
\begin{verbatim}
getChunkSize(bandwidth, rtt, eps) {
   bdp = bandwidth * rtt
   sst = bdp * (3/4)
   r_1 = rtsToSST(sst)
   r_2 = rtsFromSSTtoBDP(sst, bdp)
   totalRts = (r_1 + r_2) / eps
   return (1 - eps)* (totalRts * bdp)
}
rtsToSST(sst) {
   cwndStart = (10 * MSS)
   rounds = Log2(sst, cwndStart)
   return max(1, ceil(rounds) + 1)
}
rtsFromSSTtoBDP(sst, bdp) {
   byteToBDP = bdp - sst
   return bytesToBDP / MSS + 1
}
\end{verbatim}
\caption{Code for determining the right chunk size.}
\label{prog:size}
\end{program}

Program~\ref{prog:size} shows our algorithm for selecting the chunk size of a
video transfer, based on the derivation above. The \texttt{getChunkSize}
function takes as input an estimate of the current bandwidth and RTT, as well
as the desired $\epsilon$. 
It sets  $r = (r_1+r_2)/\epsilon$ and then uses Equation~\ref{eq:efficiency} to determine the chunk size.
The program make the following conservative assumptions since it cannot access
the internal state of the congestion control protocol: (i) that \cwnd drops to
its initial value between requests and TCP re-enters slow start, and (ii) that
SST is set to $3/4$ of fair-BDP instead of $3/4$ the maximum \cwnd as per TCP's
specification.
%~\cite{RFC:2861}.

%%It computes $r_1$ and $r_2$ using the earlier
%%equations and sets $r = (r_1+r_2)/\epsilon$ as prescribed above.
%%It then uses Equation~\ref{eq:efficiency} to determine the chunk size.
%Note that the values
%for $r_1$ and $r_2$ are conservative because they assume \cwnd  drops to its
%initial value between chunk requests and TCP re-enters slow start. The video
%player makes this pessimistic assumptions since it cannot determine the state
%of the congestion control protocol (\eg if it is in slow-start or
%additive increase).
%%, but a browser-based video player does not
%%have more detailed information about the state of TCP's congestion window. 
%% on which to make a more informed decision. 
%Likewise, the algorithm estimates the SST at $3/4$ of fair-BDP
%instead of $3/4$ the maximum \cwnd as per TCP's
%specification~\cite{RFC:2861}.
%%sets the SST to $3/4$ of the max \cwnd reached before an idle
%% period~\cite{RFC:2861}.

In our implementation, we use standard techniques to estimate RTT and
bandwidth: the RTT is obtained by measuring the latency of a 10-byte XHR
request that we send to the video server every second. Bandwidth is estimated
by taking an EWMA over the measured throughputs of video segment downloads
while downweighting smaller requests.

\ignore{
Program~\ref{prog:size} shows the algorithm for defining the chunk sizes to use
for the video transfer. The \texttt{getChunkSize} function starts with
an estimate of the current bandwidth and RTT as well as the desired $\alpha$.
It then computes $r_1$, \ie the number of
round-trips from the initial \cwnd value to the slow start threshold in
exponential increase. 
It also computes
$r_2$, the
number of round-trips from the slow start threshold until \cwnd reaches fair-BDP in
additive increase. Note that this is a conservative estimate for $r_1$ and
$r_2$ since the \cwnd will probably not drop to its initial value between
requests, but a browser-based video player does not have more detailed information
about the state of TCP's congestion window.
% on which to make a more informed decision. 
Likewise, the algorithm estimates the slow-start threshold at $3/4$ of
the flow's fair-BDP, as the TCP specification sets the SST 
to $3/4$ of the max \cwnd reached before an idle period~\cite{RFC:2861}.
}

\ignore{
Finally, we derive the chunk size by using the fair-BDP estimate to figure out how
much data will be transferred in a given amount of round-trips.  This algorithm
results in chunk sizes that are practical and that allow video flows to achieve
their fair share of network bandwidth, as we show in Section~\ref{sec:eval}.
}

We note that there is no interaction between
Program~\ref{prog:size} and the ABR algorithm choosing the bitrate.
A chunk may contain multiple segments of varying bitrates.
The data plane simply ensures that a chunk size of data is continuously
downloaded.

\subsection{Two data plane implementations}
\label{sec:dataplane}

Program~\ref{prog:size} tells us the right chunk size to use, but does not tell
us how to download this amount of data. We describe two simple data plane
mechanisms for performing the actual downloads. The first, called \system,
pipelines multiple video segment requests together to comprise a larger
(chunk-sized) download. The second, called \systemx, issues a single range
request that spans enough video data to meet the chunk size. Both players are
implemented as modifications of dash.js~\cite{dash-if}.

We allow the video player to adapt the bitrate in the middle of a
chunk, and use the minimum chunk size to ensure the video flow achieves its
fair-share throughput.  However, whereas \system achieves these goals
simultaneously, \systemx imposes a tradeoff between bitrate adaptation and
throughput.  On the flipside, \system is not
readily implementable using standard web browser APIs, and hence must be
deployed as a browser extension.

\subsubsection{\textbf{\textit{\system: Pipelined requests}}}
\label{sec:pipeline}

Our first mechanism uses HTTP pipelining to string multiple video segment 
requests together.  By structuring this pipeline carefully, we can ensure 
the video server is never waiting for an HTTP request and that
the ABR algorithm can change bitrates in a timely manner.
\footnote{In some scenarios, \eg live video streaming, pipelining is
difficult because the data is not yet available. We do not consider such
scenarios.}

%while 
%the ABR algorithm can still change the bitrate in the middle of a
%chunk.\footnote{In some scenarios, \eg live video streaming, pipelining is
%difficult because the data is not yet available. We do not consider such
%scenarios.}

%Although  each individual HTTP request is for a single video segment,
%due to pipelining, the video server does not end up waiting for the next
% segment request upon completing the previous one
%If all segment requests belonging to the same chunk are pipelined together,
% then it will ensure uninterrupted transfer of the entire chunk to the player.

We refer to the segment requests belonging to the same chunk as a
\emph{train}; the size of the train is exactly the chunk size from
Program~\ref{prog:size}.  Clearly, if we pipeline a train all at once, the ABR
algorithm will not be able to change bitrates in the middle of a chunk.  This is
a problem because the chunk size may be quite large.  Instead, we pipeline the
requests incrementally, by limiting the number of \emph{outstanding} requests
in the network.  As soon as a request completes (reducing the outstanding
requests by one), a new segment can be requested. This continues until the
entire train has been issued.

In order to set the number of outstanding requests, we
observe that we only need enough of them to ensure the video server is not idly
waiting in the middle of a chunk.  Thus, we set this number to generate 
a response of at least \bdp bytes (but enforce a minimum of 2 to avoid
sequential downloads).  Although the ABR algorithm cannot change the bitrate of
outstanding requests, in practice only a few are needed and they also complete
within a few RTTs.
%---\eg for 4-second segments and a network with RTT
%less than 12s, requesting four segments is sufficient.
% (almost all terrestrial communication).

{\bf Why not normal HTTP pipelining?} One might wonder why a minimum train size is needed if requests are being
pipelined, since the video player is continuously downloading data anyway.  The
answer is that it is needed for when the buffer fills and the player oscillates
between downloading data and pausing to drain the buffer. Some ABR
algorithms explicitly avoid filling the buffer (\eg~\cite{Huang14}), but many
do not. To serve all cases, \system enforces the minimum train size each time
downloading resumes, {\em even if this causes the buffer to overfill}---the
latter can easily be accommodated by allocating a small amount of additional
space, no larger than the minimum train
size. Our evaluation shows this is necessary for good performance
(Figure~\ref{fig:abr_universality}). We also show that the maximum buffer size
remains bounded in practice (Figure~\ref{fig:perf}), since its growth 
depends on the gap between video bitrate and available throughput, which the ABR
algorithm tackles.
%growth rate is proportional to the difference between the network throughput
% and video bitrate and the ABR algorithm adjusts the video bitrate to track
% the throughput (switching the video bitrate dynamically throughout the train).

Browsers do not currently provide sufficient APIs to HTML5 video
players to implement \system. Neither the XMLHttpRequest API 
nor the new experimental Fetch API~\cite{fetchAPI} expose control
over how requests map onto TCP connections.  This makes it impossible to
form pipeline trains in a controlled manner.  Thus, we implement \system as a
Chrome extension, which uses lower-level socket API.  In contrast, \systemx can
be implemented using standard HTML5 video.

\subsubsection{\textbf{\textit{\systemx: Expanded range requests}}}
\label{sec:solution:expanded-request}

Our second mechanism requests a larger range of data in each HTTP
request.  In \dash video, it is common for servers to store a video as a single,
contiguous file.  To fetch individual segments, players use the HTTP
Content-Range header
%~\cite{rfc2616} 
to specify a byte range within the file.
\systemx simply increases this byte range to span at least the chunk size.  Note
that this approach will not work if video segments are stored as separate files.
In addition, requesting variable segment sizes may interact poorly with caching
services such as CDNs.
%(though the latter problem is typically easy to fix).

To change the bitrate in the middle of a chunk download, the video player can
call the cancel() method on the current HTTP request and issue a new
request. Canceling a request closes the underlying TCP connection, and starting
a new connection incurs a throughput penalty.  Thus, frequent bitrate
changes will decrease overall throughput.  As shown in our evaluation, this
tradeoff disadvantages \systemx compared to \system. 

\reversemarginpar

\section{Evaluation}
\label{sec:eval}

We compare the performance of \system against leading industry players, and
answer the following questions about \system's performance in various scenarios:

\begin{itemize}[itemindent=0pt,leftmargin=25pt]
  
  \item[\S\ref{sec:network-params}] When competing
  against bulk flows across varying bottleneck bandwidths, queue sizes, and
  number of competing flows;
    
  \item[\S\ref{sec:network-params}] When competing
  against other video flows;
  
  \item[\S\ref{sec:abr-universality}] With many
  different control plane algorithms;
  
  \item[\S\ref{sec:abr-universality}] When evaluating whether the pipeline train is 
  necessary for good performance;
  
  \item[\S\ref{sec:industry}] Compared to today's video players;
  
  \item[\S\ref{sec:systemx}] Compared to the expanded-range-request approach
  of \systemx, as opposed to its own pipelining.

  %\item How does Sprint perform when the video buffer is full?

 % \item How do players that use multiple TCP flows compare to our approach? 
  
\end{itemize}

We start by establishing a baseline implementation for industry players. Then,
we compare the performance of our player to that baseline. 

\subsection{Experimental setup and methodology}

{\bf Fixed broadband networks.}  To evaluate video performance on
fixed broadband networks, we emulate a range of bottleneck network conditions. 
We connected (via wired Ethernet) two laptops to a Cisco
E1200 home router. We installed DD-WRT on the router and used Linux's
token-bucket filter (\texttt{tbf}) to adjust downstream bandwidth and queue
sizes. In all of our experiments, our ISP's actual downstream bandwidth was
greater than that permitted by the token bucket.
Unless otherwise specified, the experiments used a bottleneck bandwidth
of 3Mbps and a queue size of 256KB (although some experiments go
up to 25Mbps and 1536KB, respectively). These are representative
values: 3Mbps was chosen from the Netflix ISP Speed
Index for the US~\cite{netflix:ispindex}, and 44\% of home internet connections have a download
queue size of 256KB or greater (see \S\ref{sec:network-params}).
The TCP buffers on both laptops were tuned to avoid being limited by TCP
flow control in all the network scenarios tested.
Flow control is only relevant if the application receiving data cannot keep up with the network
or if the TCP buffers are too small for the network. We believe these issues are rare
due to the availability of high-performance browsers and TCP buffer auto-tunning.

{\bf Mobile networks.} We also performed experiments on mobile
devices running on the T-Mobile network. No traffic shaping was used for these
experiments. 
\ma{illustrating that our results are not a by-product of how we
shape bandwidth for the fixed broadband experiments.} 

{\bf Measuring performance.}
From an experimental point of there is a
very clear decomposition: the data plane is evaluated based on whether it
achieves its fair share of network bandwidth while the control plane is
evaluated on how closely the video bitrate tracks the network rate.
Thus, we report both the network throughput achieved by the video flow and QoE
measures such as video bitrates, buffer levels and number of rebuffering
events.  Though, since we propose a data plane solution that is independent of
the control plane, we concentrate on measures of network throughput. 
The video bitrate has to be slightly lower than network throughput on average in order to
have uninterrupted video playback. So any problems in network throughput
necessitate inferior video quality downloads.

% MJF Next line ``if player ... interruptions'' does not parse

We measure fairness using two metrics. For experiments where multiple video
flows compete again each other, we use the Jain's fairness index
(JFI)~\cite{jain}. For experiments where a video flow competes against a bulk
flow, we use the percent of fair-share throughput achieved by a video flow.  We
use percent fair-share instead of JFI in the latter scenario because it is
easier to interpret (if a flow achieves X\% of fair-share throughput, then the
video bitrate is at most X\% of optimal) and because it isolates
the video flows from the competing bulk flows, which we
already know achieve their fair share.

We define fair share as the total bandwidth divided by the number of flows.
We measure total bandwidth using \texttt{tshark} traces by summing the throughput
of all flows.
This allows us to measure fair-share in networks we do not
control (\eg in our cellular network experiments), and relies on the fact that
bulk flows expand to consume any unused bandwidth.

\ignore{
\hil{In order to be able to re-use the same metric in experiments where we
  control the available throughput and where we don't (\eg experiments over
  cellular networks), we use the sum of the throughput of all flows as the
measure of total bandwidth}. We measure network throughput by analyzing 
\tnote{\texttt{tshark}}{Is the trace overhead low?} traces collected during
each experiment.
}

%Our measure of fairness is easy to interpret: if a flow achieves X\% of
%fair-share throughput, then the video bitrate can be at most X\% of the optimal
%rate.   However, we use JFI when video flows compete against other
%video flows.
%
\ignore{
We choose this measure of fairness instead of alternatives like Jain's fairness
index (JFI)~\cite{jain}, as it is easier to interpret and focuses on the at-risk video flows rather than competing bulk TCP flows (which experiments
show receive their fair share).  In other words, JFI incorporates all flows in
the network, while we focus on the vulnerable video flows to better understand
their behavior.
}

%From an experimental point of there is a
%very clear decomposition: the data plane is evaluated based on whether it
%achieves its fair share of network bandwidth while the control plane is
%evaluated on how closely the video bitrate tracks the network rate. The video
%bitrate has to be slightly lower than network throughput on average in order to
%have uninterrupted video playback. So any problems in network throughput
%necessitate inferior video quality downloads.

{\bf ABR algorithm.}
Our evaluation often compares \system against the \dash player as a baseline,
since \dash performs no worse than the leading industry players.
% MJF Not sure exactly what this sentence means.
To isolate the effects of our data plane solution, both \dash and \system use
the same (control plane) ABR algorithm.
Thus, they only differ in how they download data: \dash downloads 4-second
video segments sequentially, while \system (\systemx) uses our pipelining
(expanded range request) solution with dynamic chunk sizing from
Section~\ref{sec:solution}.

We use an ABR algorithm modeled
after the best-in-class solution of Huang et al.~\cite{Huang14}, unless
otherwise specified. The algorithm selects the video bitrate based on the
level of the video buffer: every time the buffer level increases (decreases) by
10 video-seconds, the bitrate is increased (decreased). To prevent oscillation,
the algorithm never switches back to the last bitrate. We chose this algorithm
(instead of newer ones \eg Yin et al.~\cite{mpcdashabr}) because
Huang's algorithm was already optimized for environments with competing flows.
Using it thus serves to highlight the problems identified (and improvements
achieved) by our work.

In all of our experiments, the maximum video bitrate is higher than the
fair-share bandwidth.  Since the ABR algorithm we use ensures that the video
buffer never fills in this case, the weaknesses shown by \dash are not due to
the pauses discussed in prior work.

\subsection{\system utilizes its fair-share of the network across
a wide-range of scenarios.}
\label{sec:network-params}

\system is able to achieve fair-share of network throughput across a wide range
of network conditions.  In contrast, \dash, which does not benefit from our
data plane mechanisms, performs poorly in many realistic scenarios.

\begin{figure*}[t]
 \centering
\includegraphics[width=\linewidth]{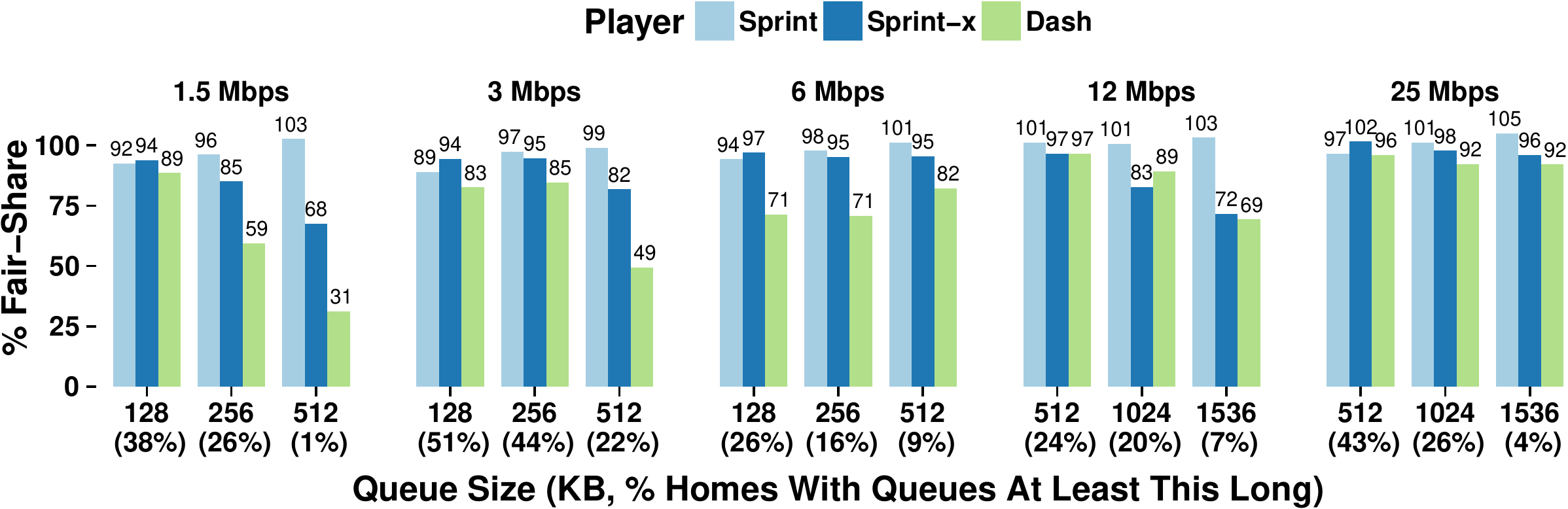}
 \caption{The median percentile fair-share of throughput used by the video players
   across five 20-minutes experiments. 
   For each bandwidth and queue size combination, we show the percentage of home internet connections
   where the download queue size $\ge$ 
   given queue size. 
 }
\label{fig:pipelining}
\end{figure*}

\begin{figure}[t]
 \centering
\includegraphics[width=\linewidth]{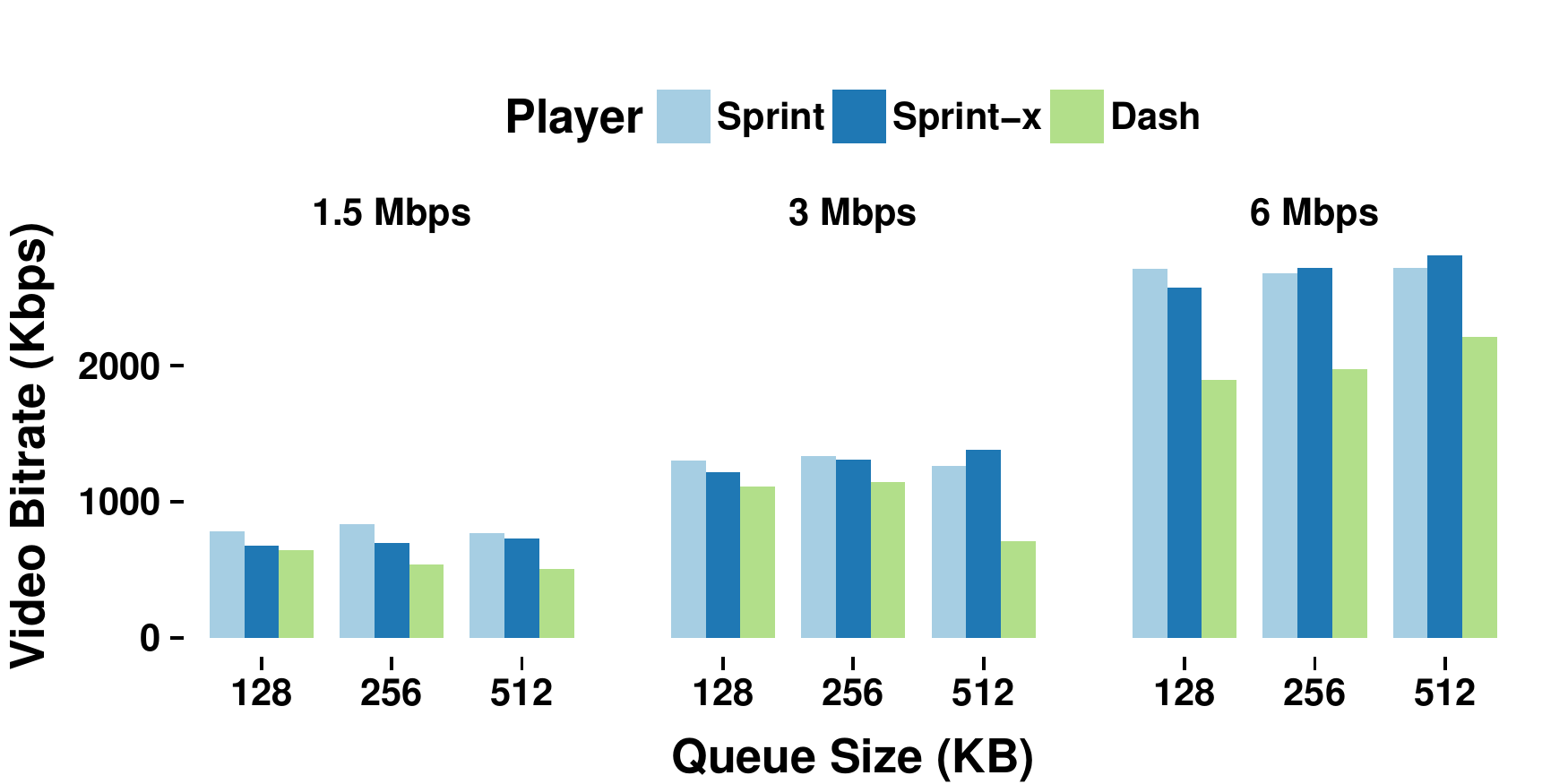}
 \caption{The median video bitrate downloaded by the video players
   across 5 experiments, each lasting 20 minutes.
 }
\label{fig:videobitrate}
\end{figure}

%\begin{table}[t]
%  \small
%  \centering
%\begin{tabular}{lcccc}
%  \toprule
%  Queue Size & 128 KB & 256 KB & 512 KB  \\
%\midrule
%\dash & 124 & 361 & 699   \\
%\system & 7 & 12 & 61 \\
%\systemx & 119 & 82 & 78 \\
%\bottomrule
%\end{tabular}
%\caption{The median number of rebuffering events experienced during video playback
%   across 5 experiments, each lasting 20 minutes for a 1.5Mbps link. At 3Mbps only the
%   DASH player experienced any rebuffering, and only at the 512KB queue size setting (108 events).
%   None of the three players experienced any rebuffering at 6Mbps.
%} 
%\label{tab:rebuffering}
%\end{table}

\begin{figure}[t]
 \centering
\includegraphics[width=\linewidth]{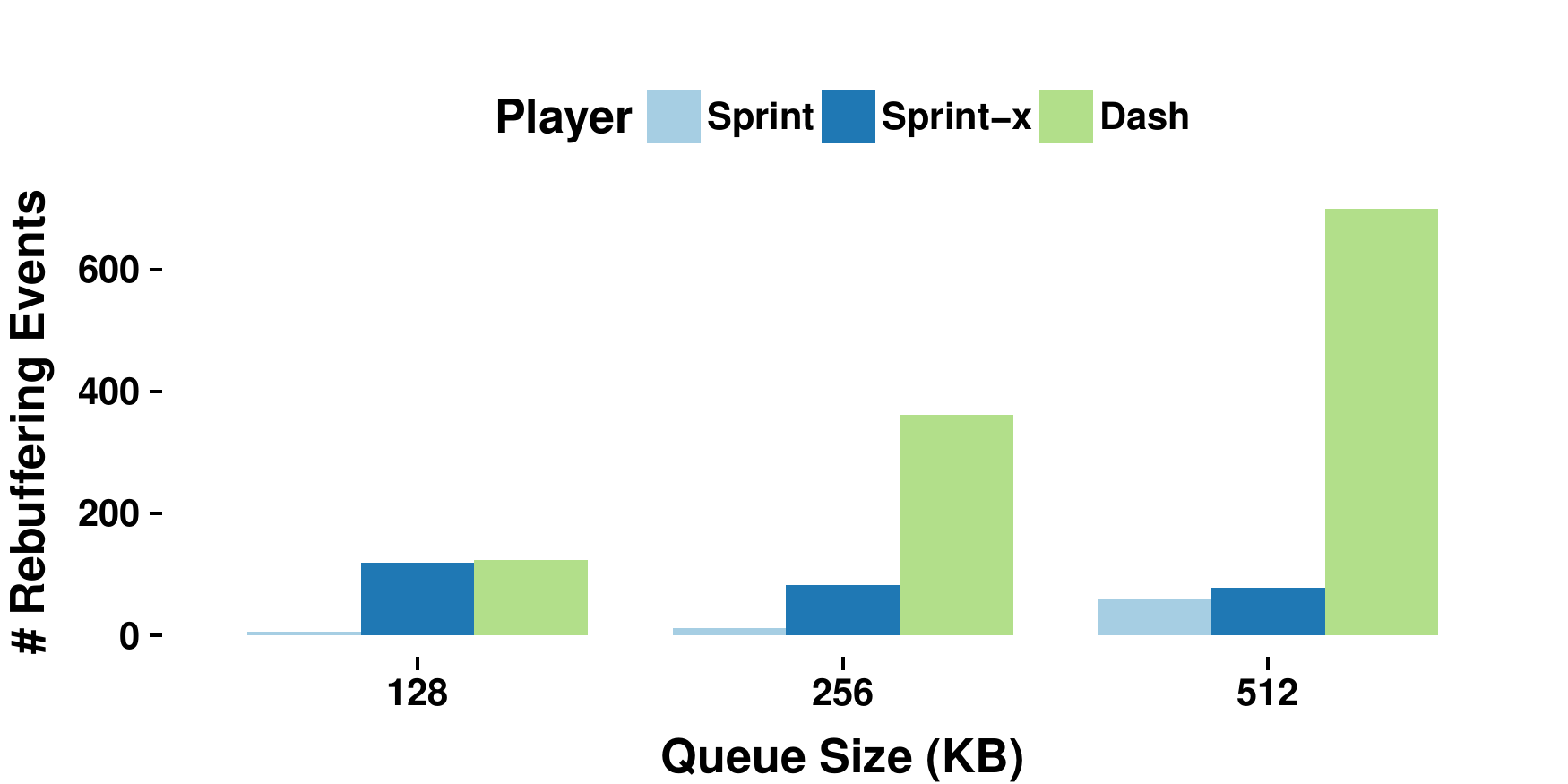}
 \caption{The median number of rebuffering events experienced during video playback
   across 5 experiments, each lasting 20 minutes for a 1.5Mbps link. At 3Mbps only the
   DASH player experienced any rebuffering, and only at the 512KB queue size setting (108 events).
   None of the three players experienced any rebuffering at 6Mbps.
 }
\label{fig:rebuffering}
\end{figure}

\paragraph{Varying bottleneck bandwidth and queue size.}
Figure~\ref{fig:pipelining} shows the percent of fair-share throughput
achieved by \system and \dash when they compete with a bulk flow downloaded
from the same server. For all values of bottleneck bandwidth and queue size tested, \system achieves its fair
share of throughput.  Our analysis from
\S\ref{sec:network} gives us the insight necessary to explain the
performance of \dash. \dash's throughput deteriorates when queue
size grows while bandwidth remains constant because competing flows induce
longer queuing delays, inflating \bdp.  Conversely, when bandwidth is increased
while queue size is held constant, video throughput improves.  In this
case, BDP does not actually increase that much because the queueing
delay decreases, but the segments are larger because of the higher
video bitrate. 

Our improvement in achieved throughput translates to a 
{\em substantial positive effect on QoE}. Figure~\ref{fig:videobitrate}
shows that both \system and \systemx are able to play higher-quality videos than \dash.
Figure~\ref{fig:rebuffering} shows that the amount of rebuffering experienced
by our players is lower.  A thorough evaluation of QoE depends on the control
plane being used, which \system tries to be agnostic to; we discuss this
further in \S\ref{sec:abr-universality}.

We use network parameters in our model that are representative of real
home networks. 
%Since our results show that performance is dependent on
%router queue size we show that the queue sizes we chose are representative.
To estimate the bottleneck queue size in these networks, we use a method similar
to the one used by Sundaresan et al.~\cite{sundaresan2011broadband}. We use
data from an ongoing study of US home networks run by the
FCC and SamKnows~\cite{fcc:study, fcc:data}.  This dataset has an
experiment in which the home router pings a server and downloads
a file simultaneously.  The maximum RTT of the pings is
representative of the bottleneck queue size.  We multiply this value by the link bandwidth
to estimate the queue size, and we use the result to determine the percent of
home connections with a certain queue size or greater.
Figure~\ref{fig:pipelining} annotates these percentages next to the
appropriate queue size and bandwidth setting.  Since performance degrades with
larger queues for players like DASH, the percentages indicate the number of
home connections whose performance will be no better than that shown.

\paragraph{Competing against other video flows.}
Table~\ref{tab:video_video} shows what happens when \dash and Sprint video
flows compete against other bulk and video flows. We evaluate (un)fairness
using the same measure as in Festive~\cite{festive}: Unfairness =
$\sqrt{1-\texttt{JFI}}$, where \texttt{JFI} is Jain's fairness
index~\cite{jain}.  Thus, a lower value implies more fairness. We show the
unfairness measure for two competing bulk flows to provide a baseline. It is
clear that \dash vs \dash performs only slightly worse than this baseline.
However, when \dash competes against a bulk flow, the bulk flow dominates and
the video flow is unable to get its fair-share of throughput. \dash is also
unable to compete with \system flows in much the same way.  In contrast,
\system, performs well when competing with other video or bulk flows.

%.  When \system competes against
%\dash, there is a lot of unfairness because \dash is unable to achieve
%its fair-share throughput in much the same way it underperforms when competing
%against bulk flows.  
%In fact, when competing against \system, \dash only
%achieves 88\% of its fair-share throughput with a queue size of 256KB, and 43\%
%with a queue size of 512KB.

\begin{table}[h]
\centering
  \small
\tabcolsep=0.11cm
\begin{tabular}{lcc  cc}
  \toprule
 Queue Size      & \multicolumn{2}{c}{256}                            & \multicolumn{2}{c}{512}         \\ 
  %\cline{2-5} 
  \cmidrule(l){2-3} 
  \cmidrule(l){4-5} 
          &  Unfairness  & Std. Dev.  & Unfairness & Std. Dev  \\
  Bulk vs Bulk & 0.03 & 0.18 & 0.05 & 0.07 \\
  \dash vs \dash & 0.04   & 0.01 &  0.10   &  0.04 \\
  \dash vs Sprint & 0.12 & 0.04 & 0.49 & 0.05 \\
  \dash vs Bulk & 0.15   & 0.02 &  0.45   &  0.02 \\
  Sprint vs Sprint & 0.06   & 0.03 &  0.05   &  0.05 \\
  Sprint vs Bulk & 0.03   & 0.02 &  0.03   &  0.04 \\
\bottomrule
\end{tabular}
\caption{Median unfairness measure (lower is better) and standard deviation 
  across 5 experiments as different types of flows compete with each other. 
  %The bottleneck bandwidth is 3Mbps.
}
\label{tab:video_video}

\end{table}

\paragraph{Varying the number of competing bulk flows.} 
When a video flow competes with multiple bulk flows, its performance is
similar to when it competes with a single bulk flow.  For example, when
competing against four bulk flows, \dash achieves a median 80\% of fair-share
throughput across five 30-minute experiments, while \system achieves
102\%.  The results for a single competing bulk flow are 85\% and 97\%, respectively.
Intuitively, adding more bulk flows reduces the video flow's fair share of
bandwidth, causing it to use a lower bitrate and thus segment size. Since both 
fair-\bdp and segment size reduce simultaneously, the net effect is
canceled out.

\begin{figure}[t]
 \centering
  \includegraphics[width=\linewidth]{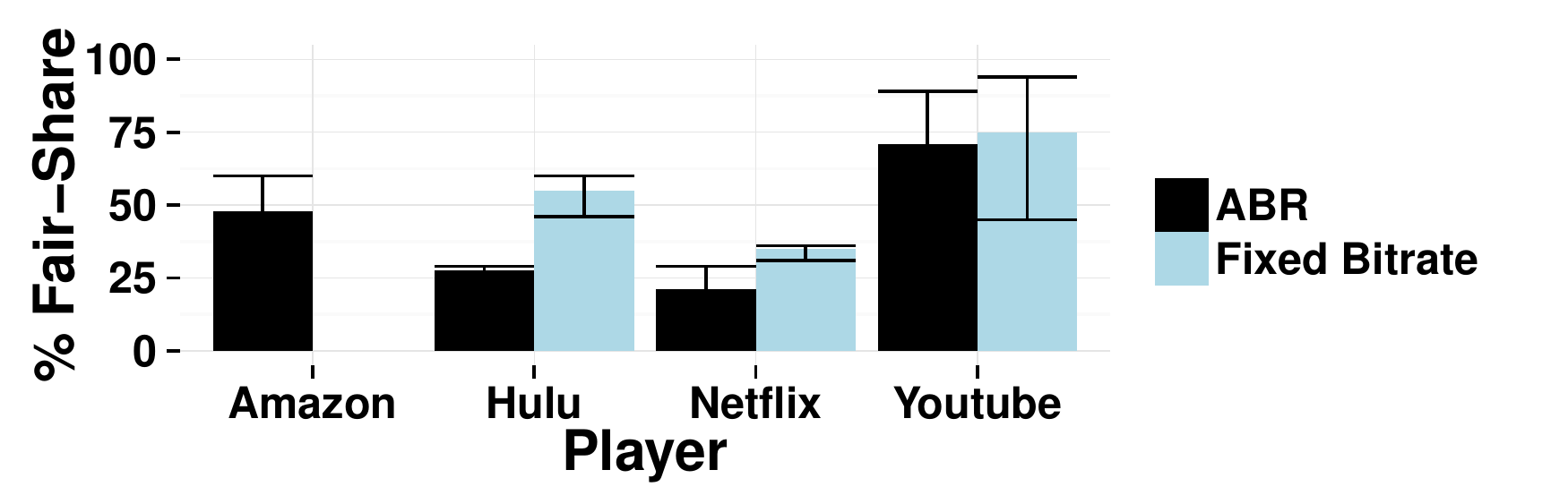} 
  \caption{The median percentile fair share achieved by industry players across
  five 25-minute experiments (error bars show max and min values). 
    For the fixed ABR experiments, the bitrate closest to the fair share of
    1.5Mbps is chosen (Amazon did not support manual bitrates).}
\label{fig:industry_bar}
\end{figure}

\begin{figure*}[t]
 \centering
 \begin{subfigure}[b]{0.32\textwidth}
  \includegraphics[width=\linewidth]{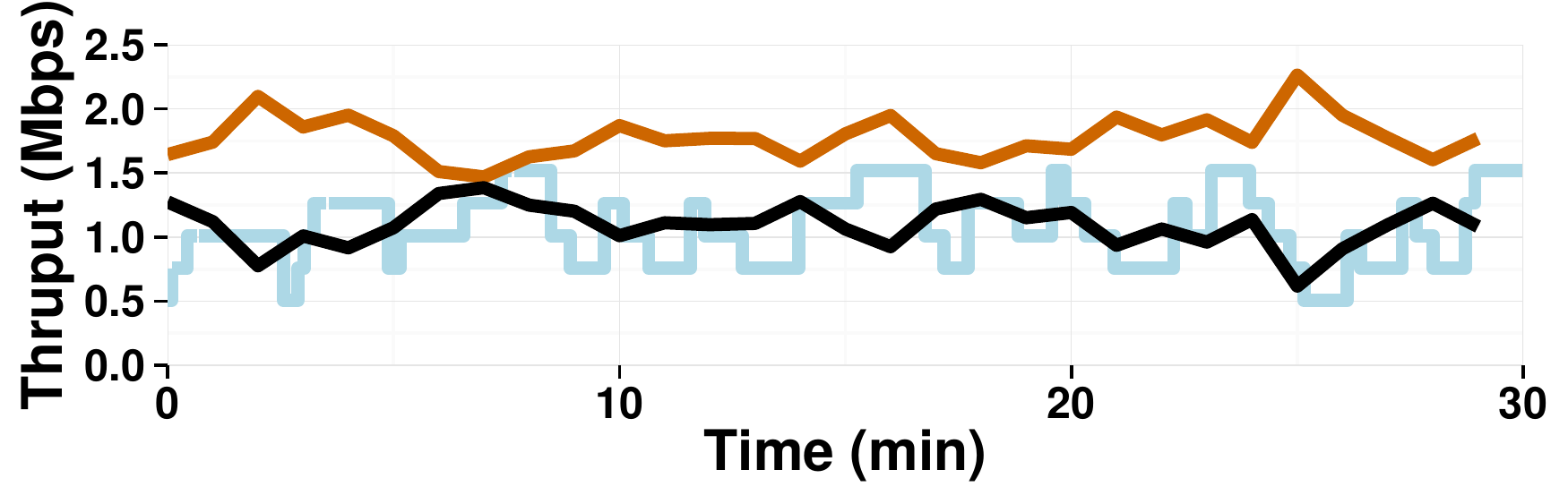} 
  \includegraphics[width=\linewidth]{./graphs/video_chromel_3_256_buffer.pdf} 
   \caption{\dash (79\% fair share) \\
   1010 Kbps median video bitrate
   }
    \label{fig:perf:dash}
 \end{subfigure}
 \begin{subfigure}[b]{0.32\textwidth}
  \includegraphics[width=\linewidth]{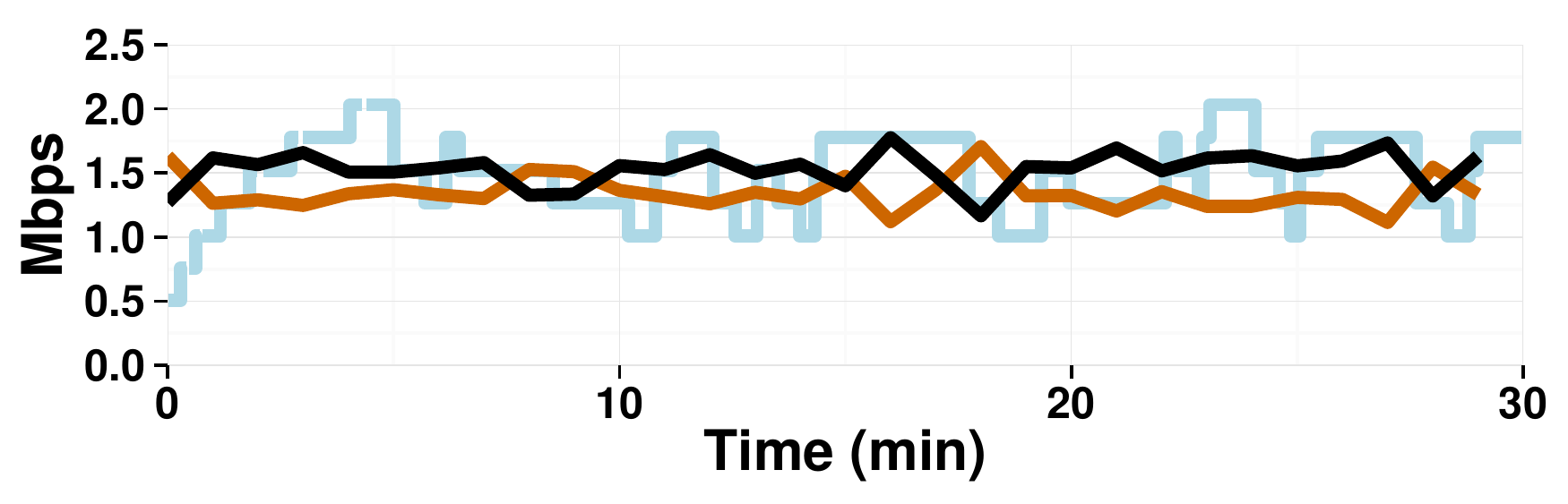} 
  \includegraphics[width=\linewidth]{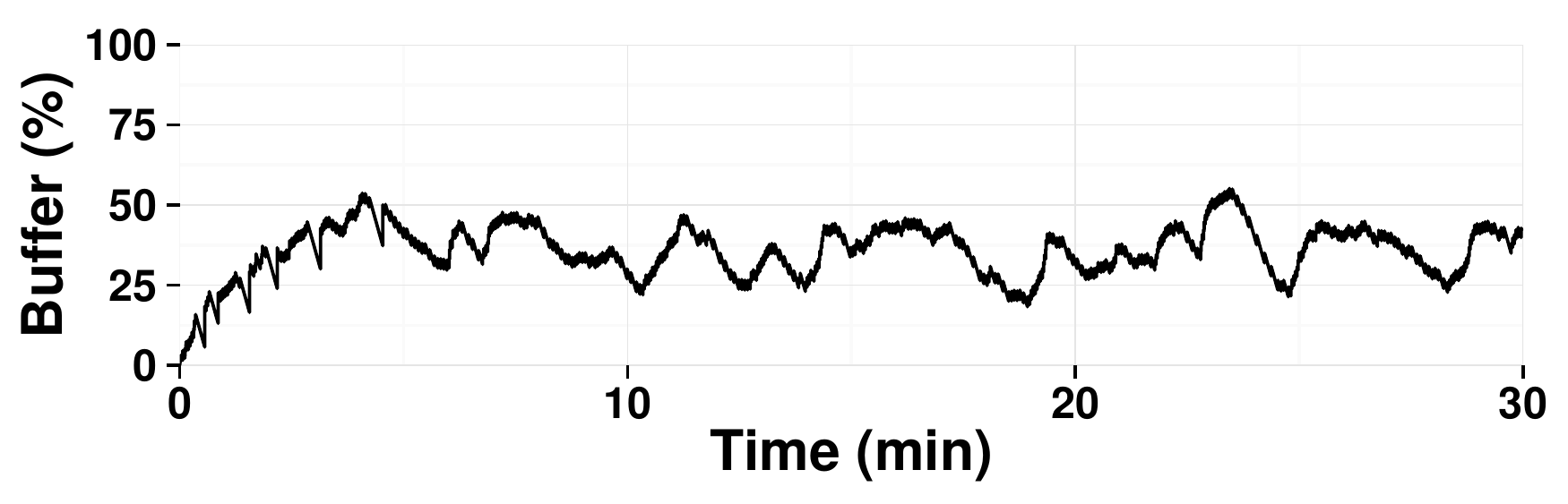} 
   \caption{\system (106\% fair share)\\
   1519 Kbps median video bitrate
   }
    \label{fig:perf:sprint}
 \end{subfigure}
 \begin{subfigure}[b]{0.32\textwidth}
  \includegraphics[width=\linewidth]{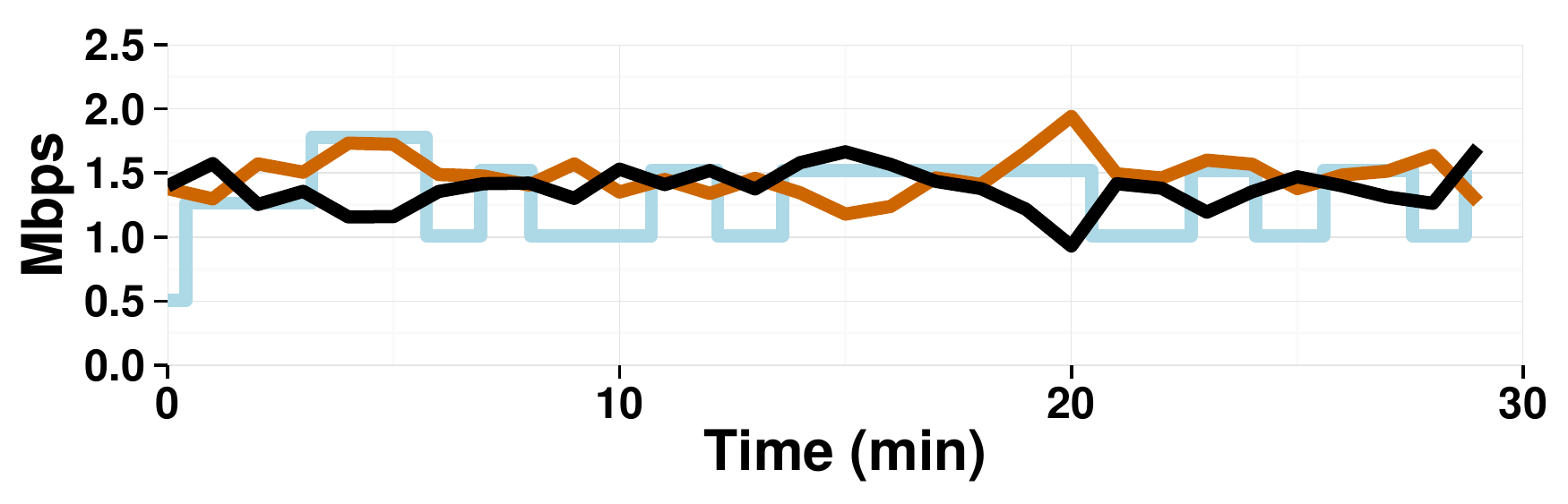} 
  \includegraphics[width=\linewidth]{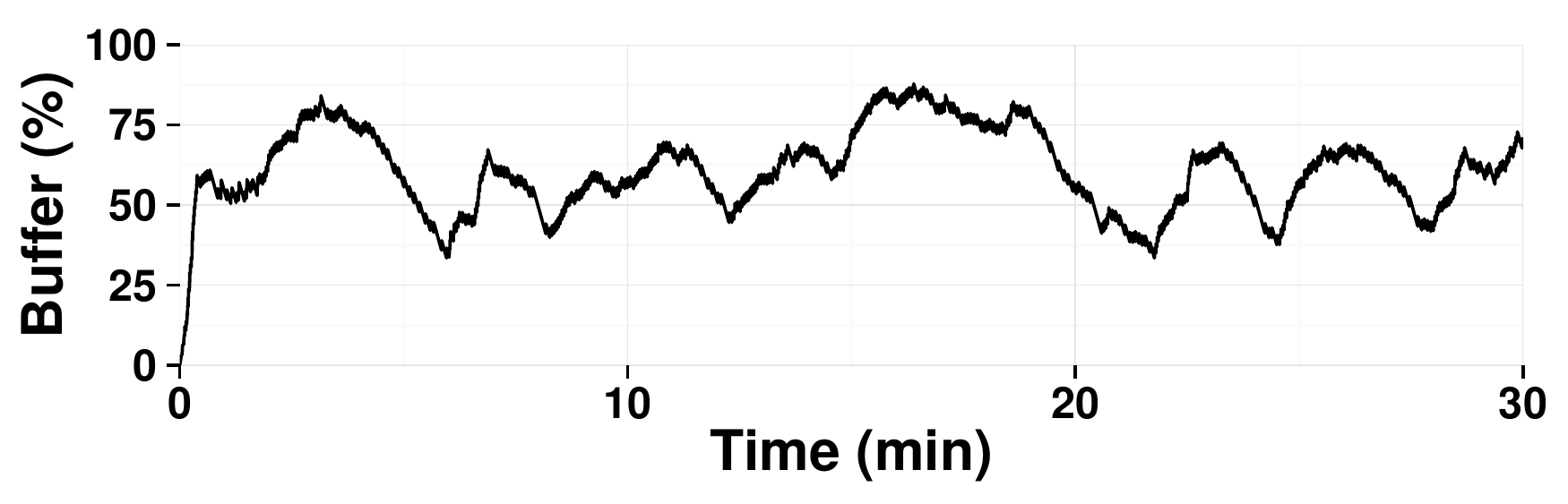} 
   \caption{\systemx (96\% fair share) \\
   1519 Kbps median video bitrate
   }
    \label{fig:perf:expanded-request}
 \end{subfigure}
 \caption{
   Aggregate throughput (black), video bitrate (blue), and video buffer levels of \dash, \system, and \systemx. In each experiment,
  the video flow shares a 3Mbps bottleneck link with a bulk file download (orange).
 }
\label{fig:perf}
\end{figure*}

\subsection{\system works well with many different ABR algorithms}
\label{sec:abr-universality}

\system provides a data plane solution that allows many different control plane
(ABR) algorithms to achieve good network performance.  This allows the control
plane to focus on optimizing the QoE without worrying about the network.

In general, ABR algorithms strive to achieve the highest possible video bitrate
without causing rebuffering.  They are often characterized by their
aggressiveness---\ie how high they make the bitrate.  To show that \system
performs well across a range of aggressiveness settings, we use a simple ABR
algorithm that matches the bitrate to a percentage of the measured throughput.
Since an aggressiveness of less than 100\% will use less than the full fair
share, we present the results in terms of {\em expected} fair share, which we
define as $\min(100\%, \text{aggressiveness})$ times the fair share.

\begin{figure}[t]
 \centering
\includegraphics[width=\linewidth]{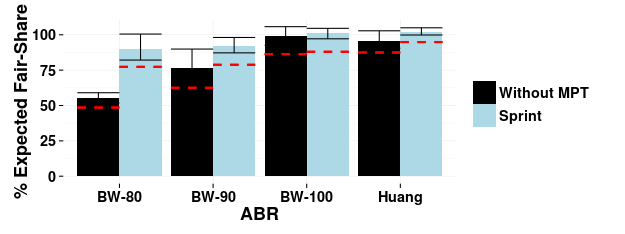}
 \caption{Median \% expected fair share achieved by \system while using
   different ABR algorithms across five 25-minute runs (error bars show min and max).
   This is compared to \system without the minimum pipeline train (MPT).
   The BW-x ABR algorithms set the video bitrate to x\% of the
   measured throughput. The red dashed line shows the video bitrate as a percentage
   of expected.
   %If x
   %is less than 100, the network fair share is weighted by x to obtain the
   %expected fair-share.
   }
\label{fig:abr_universality}
\end{figure}

As Figure~\ref{fig:abr_universality} shows, \system achieves its fair share of throughput even when
the ABR algorithm is less aggressive. This is a challenging scenario
because the video buffer often fills and causes pauses in downloading.
Prior work showed that video players underperform in this
regime~\cite{Akhshabi12, Huang12} and proposed new ABR algorithms to
avoid pauses~\cite{festive, Huang14}.  In contrast, \system uses
the pipeline train size to ensure that enough data is transferred between pauses
(\S\ref{sec:pipeline}). Figure~\ref{fig:abr_universality} shows that when
the train size is not enforced, network throughput degrades significantly for
aggressiveness settings less than 100\%.  Many ABR algorithms fall into this
category; for example, by default DASH has an aggressiveness of 90\%.

%the DASH's default ABR algorithm that ships with DASH has an aggressiveness of 90\% .
%due to the desire to avoid rebuffering events, which negatively impact QoE.  

The interaction between \system and the ABR algorithm is complex.  In some
cases, \system may obviate the need for a control-plane feature (\eg
aggressiveness); or vice versa, \eg Huang et al's~\cite{Huang14} algorithm
avoids filling the buffer, making \system's minimum train size unnecessary.

\begin{figure*}[t]
 \centering
  \includegraphics[width=\linewidth]{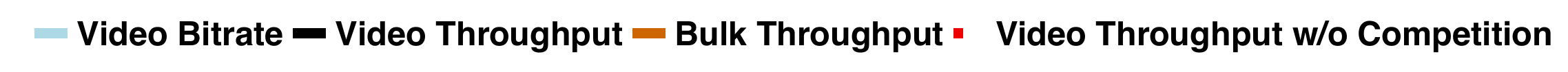} 
 
  \begin{subfigure}[b]{0.24\textwidth}
    \includegraphics[width=\linewidth]{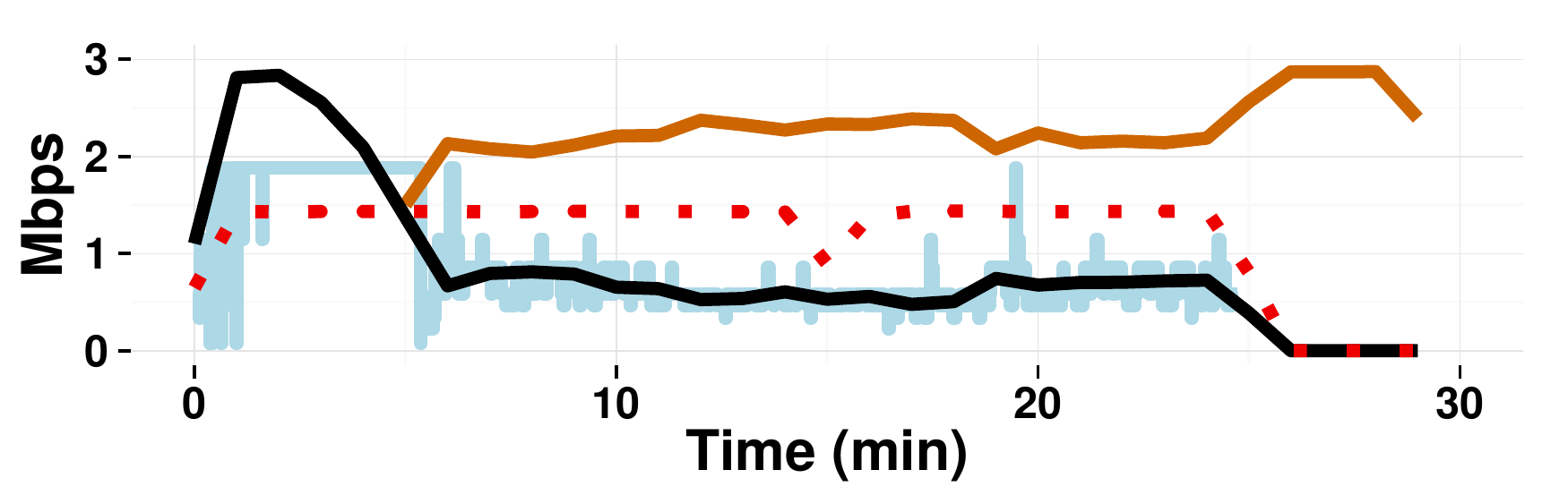}
    %\captionsetup{justification=centering}
    \caption{Amazon (48\% fair-share) \\
    583 Kbps median video bitrate}
    \label{fig:amazon_app}
 \end{subfigure}
  \begin{subfigure}[b]{0.24\textwidth}
    \includegraphics[width=\linewidth]{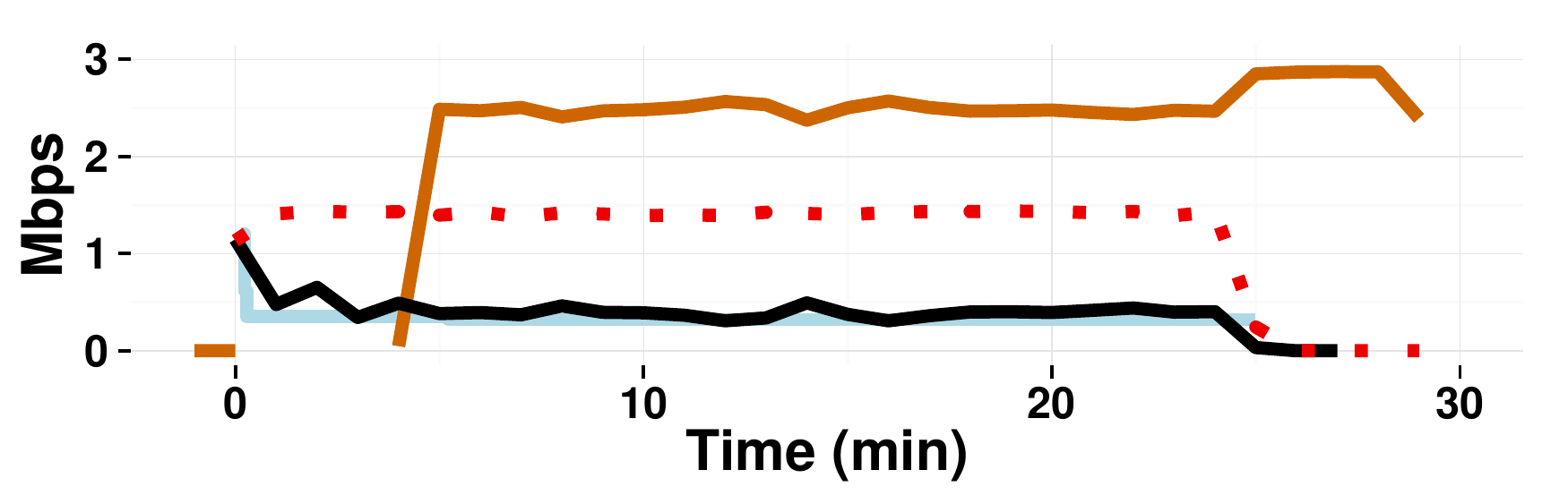}
    %\captionsetup{justification=centering}
    \caption{Hulu (27\% fair-share) \\
      320 Kbps median video bitrate
  }
    \label{fig:amazon_app}
 \end{subfigure}
  \begin{subfigure}[b]{0.24\textwidth}
    \includegraphics[width=\linewidth]{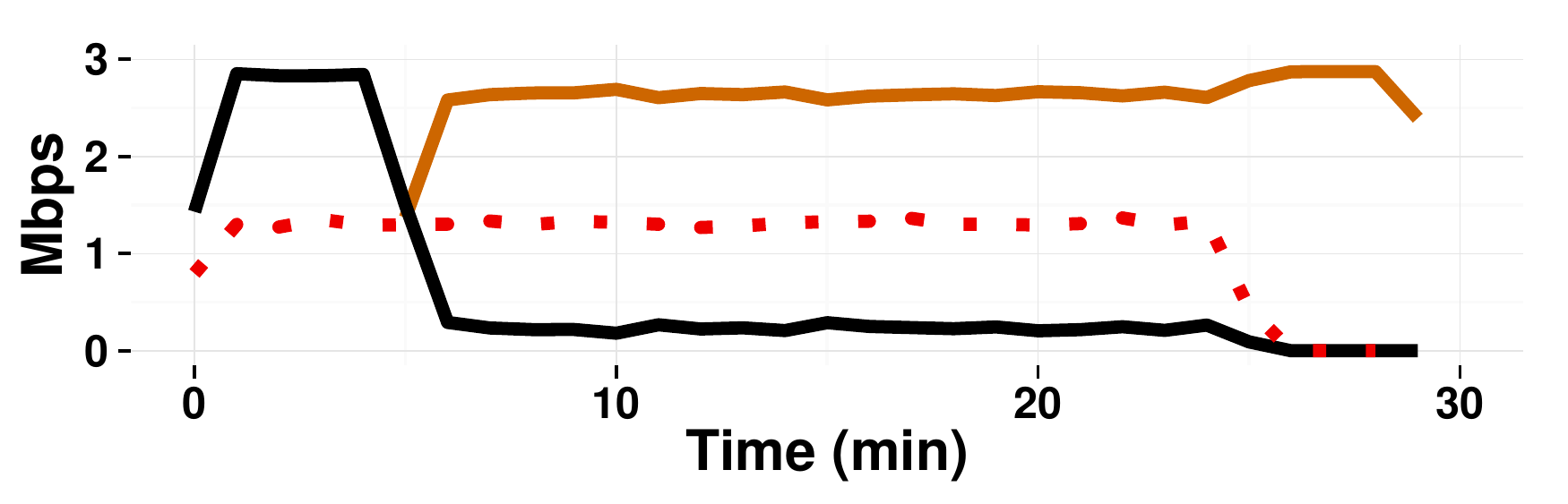}
    %\captionsetup{justification=centering}
    \caption{Netflix (21\% fair-share)\\
   Video bitrate n/a }
    \label{fig:amazon_app}
 \end{subfigure}
  \begin{subfigure}[b]{0.24\textwidth}
    \includegraphics[width=\linewidth]{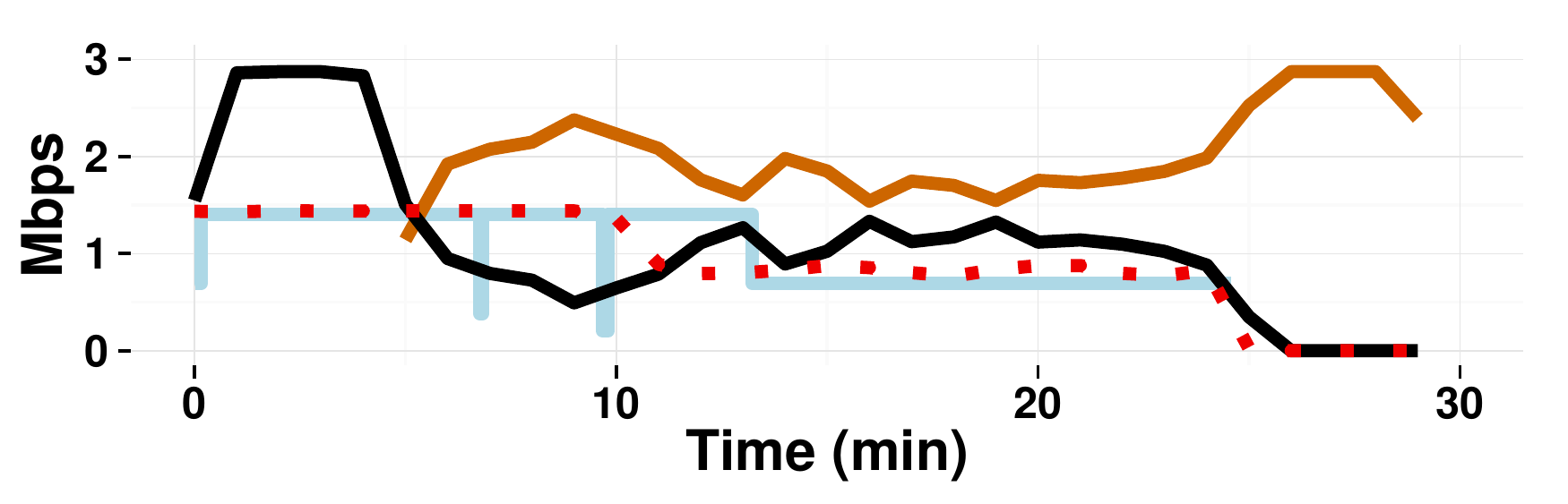}
    %\captionsetup{justification=centering}
    \caption{Youtube (71\% fair-share) \\
                691 Kbps median video bitrate}
    \label{fig:youtube_app}
 \end{subfigure}
 \caption{Aggregate throughput of various industry players when
 sharing a bottleneck link with a file download.  Black lines are the
 minute-averaged network throughput of the video players; orange lines
 represent the bulk file download.  Blue lines show the bitrate of the video being downloaded.
 During the first (resp. last) 5 minutes, the only
 active flow is the video (resp. file) download; thus \% fair share and median video bitrate is calculated
 for minutes 5-25. The dotted line shows the video
 throughput of a separate experiment which is run over a 1.5Mbps link with no competing flow. The video
 bitrate is derived from the URL of the file downloaded; it could not be determined for Netflix.
 }
\label{fig:compare_app}
\end{figure*}

\subsection{\system outperforms existing players}
\label{sec:industry}

% MJF Is there a reason you choose the order you do?  You keep changing order
% of YT, Amazon, Hulu, Netflix.
%

To demonstrate that even leading industry video-on-demand services fail to
achieve their fair share of throughput, we evaluate the performance of
Youtube, Netflix, Hulu, and Amazon Video. For each service, we stream videos via a web
browser\footnote{We use Chrome for Netflix, Youtube, and Amazon Video. We use Firefox
for Hulu, the only browser under Linux on which Hulu runs.} while simultaneously
downloading a large file through the browser.  Both the video and file download
flows share the bottleneck link created by our home router or mobile connection.
The video streams are unmodified and thus incorporate all of the
services' network and CDN optimizations.

% MJF is there a reason the figure is so far behind citation?  Annoying!
%
% For amazon, what does it mean when video goes to 0 at the end?!?!  Same with
% netflix.  Did the video finish downloading beforehand?  Might want to cut
% that data if so...makes it seem like the services can't get any data to
% download, when in fact nothing to download.

\paragraph{Fixed broadband networks.}  Figure~\ref{fig:compare_app} traces the
aggregate throughput of the video and file download flows for Netflix, Amazon,
Hulu, and Youtube. The large gap in throughput that develops between the two
flows indicates that the video flow is unable to achieve its fair share when
competing against bulk flows. At the same time, the dotted line shows that 
in the absence of competition, these players are able to use their fair share to
achieve a higher video bitrate.

We now show that it is the data plane, not the control plane, that is mostly 
responsible for this gap.  The control plane can negatively affect a video
flow's throughput if it stops requesting data or inserts pauses between
requests, such as when the video buffer fills.  Conversely, if the control
plane is continuously requesting data, then the data plane should in principle
be able to achieve the full fair share of throughput.  We measured the pauses
between consecutive requests for all players and found that they were less than 100ms at
the 95th percentile for Amazon, Netflix, and Youtube. This suggests that the
control plane was not the culprit for these players.  Hulu, on the other hand,
had significant pauses between requests: 1373ms at the 95th percentile.  

To obtain a definitive answer, we conducted a controlled experiment that forced
each player to use a fixed video bitrate close to the fair-share throughput,
thus bypassing the ABR algorithm.  All players except Amazon provide a
setting to do this.  Figure~\ref{fig:industry_bar} shows the results. 
Using the fixed bitrate, the pauses between requests for Hulu were reduced to
128ms at the 95th percentile, and in general the video buffer never filled for
any of the players.  For all players, the degradation in throughput caused by
using the ABR algorithm is only a small part of the total degradation from
100\%. This means that the remaining degradation is due to the data plane.

%In each of these experiments the video bitrate is set the value closest the the
%network fair-share and the measured throughput is always less than the video
% bitrate.

Youtube does not perform as well as the other players when operating without
competing flows at (Figure~\ref{fig:compare_app}d). This is due to a large gap
in bitrate encoding levels at this bandwidth setting: the closest bitrates
are 727Kbps and 1469Kbps. Youtube's ABR tends to drop to the lower
bitrate in the middle of playback. For the fixed bitrate experiment in
Figure~\ref{fig:industry_bar}, we forced the player to use the higher bitrate
throughout, but Youtube still used only 75\% of its fair-share.

Using the same experimental setup, we evaluate our \system and
\systemx players as well as the \dash player. Figure~\ref{fig:perf} clearly shows that both
\system and \systemx outperform \dash and achieve close to---slightly more,
in the case of \system---their fair share of throughput. Similarly, our players
achieve a higher median video bitrate. The buffer level graphs show that none of
the players experience rebuffering.

\comment{NEED TO FIX ORDER OF FIGURE APPEARANCE THROUGHOUT. Also, too much
reliance on long figure captions.}

%\begin{table}[h]
%\begin{tabular}{llllll}
%        & \multicolumn{3}{c}{App}                            & \multicolumn{2}{l}{Tether}         \\ \cline{2-6} 
%        & \% Fair-Share & Competing Kbps & No-Competing Kbps & \% Fair Share & Non-Competing Kbps \\
%Youtube & 49\%          & 543            & 2004              & 58\%          & 2148               \\
%Netflix & 27\%          & 530            & 881               & 53\%          & 3031               \\
%Hulu    & 18\%          & 351            & 690               & 64\%          & 1483               \\
%Amazon  & 11\%          & 288            & 1358              & 20\%          & 2544              
%\end{tabular}
%\end{table}

\ignore{
\begin{table}[t]
  \small
\begin{tabular}{lcc  cc}
  \toprule
        & \multicolumn{2}{c}{App}                            & \multicolumn{2}{c}{Tether}         \\ 
  %\cline{2-5} 
  \cmidrule(l){2-3} 
  \cmidrule(l){4-5} 
  & \% Fair-Share             & \% Fair Share           \\
  Youtube & 49\%              & 47\%            \\
  Netflix & 27\%              & 48\%           \\
  Hulu    & 18\%              & N/A            \\
  Amazon  & 11\%              & 30\%              \\
\bottomrule
\end{tabular}
\caption{Percent of network fair-share used by video services run over cellular networks.
  The video flow competes against a single file download; the measurement period lasts 10 minutes. 
  Hulu could not sustain playback while using a tethered connection.
}
\label{tab:industry_mobile}
\end{table}
}

\paragraph{Mobile networks.} We tested the mobile performance of the industry
players by using each service's mobile app, and also by running the
web browser version of the service on a laptop connected to the
Internet via USB tethering. Both experiments were instructive.  The app
experiment revealed that these services did not implement any (effective)
special logic to compensate for the large latencies of mobile networks. At
the same time, the tethering experiment made sure that the performance of the
video player was not affected by any limitations built into the app
(\eg to save mobile data).  The tethering experiment was also necessary because
we noticed that the throughput of some of the apps was limited by the TCP
send window, perhaps due to limited receive buffers. The tethering
experiment ran on a laptop with a TCP stack tuned so as to avoid this
limitation.

Table~\ref{tab:industry_mobile} shows the results when playing a video 
concurrently with a file download. Both the app and
tethering experiments show that the video services are not able to achieve their
fair share of throughput over mobile networks.

\note{Explain upfront why you need two setups for mobile tests. Also, explain
why you need multiple competing bulk flows. FIXED?}

In contrast, \system is able to fully utilize its fair share of mobile
bandwidth.  We tested the \system and \dash players by streaming a
video for 10 minutes while performing a file download on a laptop connected
by USB tethering.
The average percent of fair-share throughput achieved over five experiments was 98\% for \system
and \tnote{only 80\% for \dash.}{You say ``only'', but this seems high compared to commercial
solutions. Why does \dash do better than Youtube et al.?}

\begin{table}[t]
  \small
  \centering
\begin{tabular}{lcccc}
  \toprule
 & Youtube & Netflix & Hulu & Amazon \\
\midrule
App & 49\% & 27\% & 18\% & 11\%  \\
Tether & 47\% & 48\% & N/A & 30\% \\
\bottomrule
\end{tabular}
\caption{Network fair share used by video services run over a mobile network
while competing against a file download. The measurement period lasts 10 minutes.
 Hulu could not sustain playback while using a tethered connection.
} 
\label{tab:industry_mobile}
\end{table}

\subsection{\systemx is not as good as \system}
\label{sec:systemx}

\begin{figure}[t]
 \centering
\includegraphics[width=\linewidth]{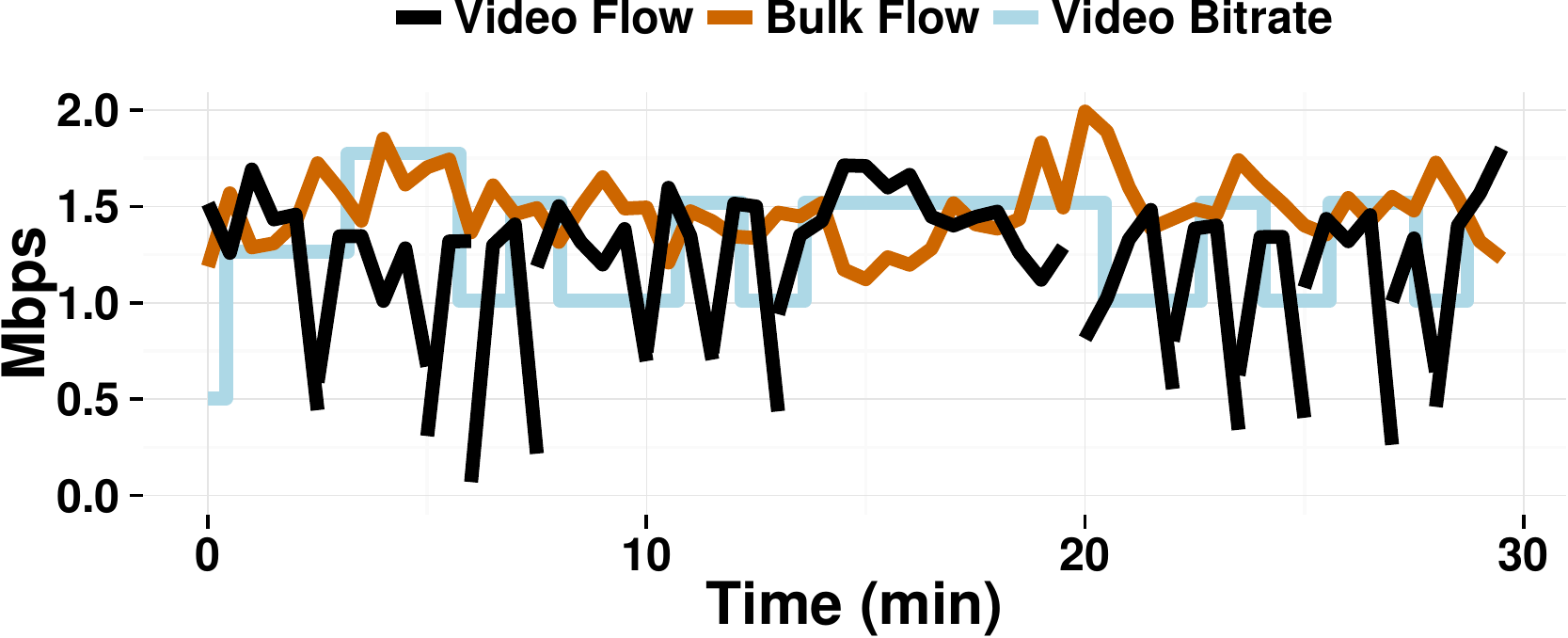}
\caption{ The throughput and video bitrates of the \systemx competing
  against a bulk flow. When the ABR changes the video bitrate, the player
  often has to close an existing flow and start a new one (each new flow
  is shown as a separate black line). This causes the throughput
  to drop.}
\label{fig:big_chunk}
\end{figure}

Figure~\ref{fig:pipelining} shows that while \systemx performs better than
unmodified \dash, it performs worse than \system. This is because \systemx has
to cancel ongoing requests when switching bitrates (as discussed in
\S\ref{sec:solution:expanded-request}). Canceling requests incurs
a throughput penalty as illustrated in Figure~\ref{fig:big_chunk}, where
the only time when the video flow achieves its fair-share is 
during minutes 15-20 when no bitrate changes occur.

%\note{Is there a buffer level stability advantage of pipelining vs. expanded
%requesets (Figure 9)? If so, could point out.} 
%as discussed in section~\ref{sec:solution:expanded-request}, the
%expanded-range-request approach used by \systemx requires the video player
%to cancel ongoing requests when switching the video bitrate. 
%%But canceling
%%a request requires closing an ongoing TCP flow and creating a new one, causing a
%%throughput penalty.  
%This is illustrated by Figure~\ref{fig:big_chunk}, which
%shows a decrease in throughput every time the ABR algorithm switches the
%bitrate. 
%%In fact, the only time when the video flow competes fairly with the
%%file download is during minutes 15-20 when no bitrate changes occur.
%Figure~\ref{fig:pipelining} shows that while \systemx performs better than
%unmodified \dash, it performs worse than \system because \system does not
%incur a throughput penalty when switching bitrates.

\section{Conclusion}
\label{sec:conclusion}

TCP dynamics interact poorly with data transfers that use small,
sequential web requests, instead of the bulk transfers for which TCP
was designed.  To enable video flows to achieve their fair-share
throughput, we derive and implement a data-plane mechanism right-sizing
requests so that these flows acts more like bulk transfers.
Our evaluation shows significant and
consistent improvements over state-of-the-art and industry video
players, with benefits across all tested ABR algorithms.

% MJF I don't get/like everythign below here.
\ignore{
But HTTP-based REST APIs for other types of data transfer are ubiquitous. The
main takeaway from this work is that, unfortunately, such data transfers cannot automatically rely on TCP to
guarantee them fair use of network resources. We leave it to future work to
explore the impact of data-plane concerns on other data-intensive HTTP-based
applications. 
}

Our approach solves the throughput problems with on-demand video streaming but
similar problems exist in real-time communication (RTC). RTC is sensitive to
increased latency and thus filling up network queues to gain throughput---as we
do in our proposed solution---is not an attractive approach in the presence of
bufferbloat~\cite{bufferbloat}. In fact there might not be any good end-to-end
solutions for mitigating latency when RTC flows compete against bulk flows in a
bufferbloated network. Instead, in-network changes to limit queue latency (\eg
fair-queing, AQM) are necessary. Such changes would also be an alternative--if
harder to deploy--solution to the video throughput problems discussed in this
paper.
%

%\ifthenelse{\equal{\ANONYMOUS}{no}}{
%\vspace{1ex}
%\subsection*{Acknowledgments}
%\vspace{1ex}

{%\small
\setlength{\bibsep}{1pt}
\bibliographystyle{abbrvnat}
\bibliography{video}
}

%\section*{Unplaced fragments, to be cut or used}
%\input{unplaced.tex}

\end{document}